\documentclass[vecphys]{svmult}

\usepackage{graphicx}        
\usepackage{bm}
\usepackage{multicol}        
\usepackage{amsmath}
\usepackage{amssymb}
\usepackage{latexsym}
\usepackage{cite}            
\usepackage[bottom]{footmisc}

\begin{document}

\title{Exciton Spin Dynamics in Semiconductor Quantum Wells}
\author{Thierry Amand \and Xavier Marie}
\institute{Laboratoire de Physique et Chimie des Nano-Objets,
UMR INSA-CNRS-UPS,
INSA, D\'epartement de Physique,
135 avenue de Rangueil,
31077 Toulouse cedex 4, France; 
\texttt{marie@insa-toulouse.fr},
\texttt{amand@insa-toulouse.fr}}
\setcounter{chapter}{0}
\maketitle

\section{Two-Dimensional exciton fine structure}
\label{sec:1}

The spin properties of excitons in nanostructures are determined by their fine structure. Before analysing  the exciton spin dynamics, we give first a brief description of the exciton states in quantum wells. We will mainly focus in this review on GaAs or InGaAs quantum well which are model systems. For more details, the reader is referred to the reviews in ref. \cite{ivchenko, Andreani}. As in bulk material, exciton states in II-VI and III-V quantum wells (QW) correspond to bound states between valence band holes and conduction band electrons. As will be seen later, exciton states are shallow two-particle states rather close to the nanostructure gap, \textit{i.e.} their spatial extension is relatively large with respect to the crystal lattice, so that the envelope function approximation can be used to describe these states. 

The problem of exciton states in bulk crystals or nanostructures is in fact a $N-$electron problem, in which we seek for the stationary states of a crystal where one electron has been removed from the valence states and set in the conduction band, thus leaving $N-1$ valence band electrons. Due to electron indiscernability, the latter states should be antisymmetrized. At this point it is more convenient to use the electron-hole pair states basis,  
 $\psi_{s,\bm k_e}(\bm r_e)$$\psi^h_{m,\bm k_h}(\bm r_h)$
where 
 $\psi_{s,\bm k_e}(\bm r_e)=\frac{1}{\sqrt V}e^{i\bm k_e .\bm r_e}u_{s,\bm k_e}(\bm r_e)$
is a conduction Bloch function ($V$ being the crystal volume), and  
 $\psi_{m,\bm k_h}^h(\bm r_h)=\frac{1}{\sqrt V}e^{i\bm k_h .\bm r_h}u_{m,\bm k_h}(\bm r_h)$
is the hole function obtained by applying the time reversal operator $\hat{K}$ to a corresponding electron valence state $\psi_{m_v,\bm k_v}(\bm r)$. This description offers the advantage to give the possibility to solve the exciton problem as a two-body problem. This is done usually in two steps: first treat the Hartree type problem between a conduction electron and a hole with direct Coulomb attractive potential, thus yielding the so-called "mechanical" exciton, then solve by perturbation the corrections due to electron-hole exchange terms. Note these terms appear due to the non vanishing coulomb exchange terms arising between the conduction electron and the remaining $N-1$ valence band electrons.  A description of the exciton fine structure in bulk semiconductors can be found in \cite{optical}.

In quantum wells structures, as in bulk material, a conduction electron and a valence hole can bind into an exciton, due to the coulomb attraction. However, the exciton states are strongly modified due to confinement of the carriers in one direction. As we have seen, this confinement leads to the quantization the single electron and hole states into subbands \cite{ivchenko,Bastard2}, and to the splitting of the heavy- and light-hole band states. The description of excitons is obtained, through the envelope function approach, and the fine exciton structure is then deduced by a perturbation calculation performed on the bound electron-hole states without electron-hole exchange. However, this approach becomes then more complex in the context of two dimensional structures, and is summarized in Appendix I. The full electron-hole wave function can finally be approximated by: 
\begin{eqnarray}
\Psi_{\alpha}(\bm r_e,\bm r_h)=\chi_{c,\nu_e}(z_e)\chi_{j,\nu_h}(z_h)\frac{e^{i\bm K_\bot \bm . \bm R_\bot}}{\sqrt A}\phi^{2D}_{j,nl}(\bm r_\bot)u_s(\bm r_e)u_{m_h}(\bm r_h)
	\label{equation1}
\end{eqnarray} 
where, $\alpha$ represents the full set of quantum indexes characterizing the exciton quantum state, \textit{e.g.} explicitly:  
$\left|\alpha\right.\left.\right\rangle = \left|s,m_h;\nu_e,\nu_h,\bm K_\bot,j,n,l \right.\left.\right\rangle$. 
Here $\chi_e(z)$ and $\chi_{jh}(z)$ are the single particle envelope functions describing the electron, heavy-hole (\textit{j}=\textit{h}) or light hole (\textit{j} = \textit{l}) motion along the \textit{Oz} growth axis, $\bm R_{\bot}$ is the exciton center of mass position, $A$ is the quantum well quantization area and $\phi^{2D}_{jnl}(\bm r_\bot) $ characterizes the electron-hole relative motion in the QW plane. This is in fact the function basis we shall take to formulate the electron-hole exchange in a QW exciton. 

The principle of the calculation relies on the evaluation of the direct and exchange integrals:
\begin{subequations}
\begin{equation}
	D_{\beta,\alpha}=\mathop{\int}_{structure}\Psi^*_\beta (\bm r_e,\bm r_h) \frac{e^2}{\epsilon_b\left|\bm r_e-\bm r_h\right|}\Psi_\alpha(\bm r_e,\bm r_h)
	\label{equation2a}
	\end{equation}
\begin{equation}
	-E_{\beta,\alpha}=-\mathop{\int}_{structure}\Psi^*_\beta (\bm r_e,\bm r_h) \frac{e^2}{\epsilon_b\left|\bm r_e-\bm r_h\right|}\Psi_\alpha(\bm r_h,\bm r_e)
	\label{equation2b}	
\end{equation}
\label{equation2}
\end{subequations} 
In the calculations of integrals (\ref{equation2}), two contributions appear: a short range one, which corresponds to the case where the electron and the hole are in the same Wigner cell $\Omega$ in the structure, and a long range one, which corresponds to the case where they are not. In the latter contribution, only the exchange integral has to be taken into account, since the direct long range Coulomb interaction has already been considered in the equations of the 2D mechanical exciton (A\ref{equationAI2}). Such integrals have been computed in ref. \cite{mas}. It turns out that, in narrow QWs, they are much smaller than the heavy-/light-hole splitting $\Delta_{hl}$, as well as the one between the different single particle subband states $\nu_{e(h)}$, and finally the 1s/2s exciton splitting. Then the first order perturbation theory, applied to the degenerated exciton states associated with a given subband pair, allows us to evaluate the corrections brought by (\ref{equation2}) perturbation.

\subsection{Short-range electron-hole exchange}
\label{sec:11}
For the ground state of the heavy-hole exciton (XH), the short range perturbation matrix is :
\begin{eqnarray}
	H^{2D(SR)}= D^{2D(SR)}-E^{2D(SR)}
	\label{equation3}
\end{eqnarray} 
It turns out that $H^{2D(SR)}$ is proportional to $\left|\phi^{2D}_{hh,1s}(r=0)\right|^2I_{hh}$, where $I_{hh}=\int^{+\infty}_{-\infty}\left|\chi_{e,1}(z)\right|^2\left|\chi_{h,1}(z)\right|^2 dz$, a measure of the probability for the electron and the hole to be at the same position in the QW, times the difference between direct and exchange Coulomb terms built with Bloch state products $\left|s,m_h\right\rangle$.
It is convenient to evaluate (\ref{equation3}) with respect to the 3D case. Then, the short range exchange is given, in the spherical approximation and within an inessential energy constant, by: $H^{3D(SR)}=\frac{1}{2}\Delta_0\bm J_e \bm. \bm S_e$, where $\Delta_0=\Omega\left|\phi^{3D}_{1s}(r=0)\right|^2$ is the 3D short range exchange splitting (in crystals of \textit{$T_d$} symmetry, an additional term $\Delta_2 \sum_{\lambda=x,y,z}S_{e,\lambda}J^3_{h,\lambda}$ introduces a small splitting in the $J=2$ exciton states, which will be neglected here, see Pikus \textit{et al.} in \cite {optical}), and $\phi^{3D}_{1s}$ is the 3D exciton hydrogenic 1s function. 

In 2D systems, due to the splitting $\Delta_{lh}$ between heavy-hole and light-hole excitons (labeled XH and XL respectively), it is possible to use the restriction of $H^{2D(SR)}$ to the XH subspace. The XH basis basis states are labeled according to their projection to the quantization axis \textit{Oz} (the structure growth axis), according to $\left|M\right\rangle=\left|s_e+j_h\right\rangle$ ($s_e=\pm1/2, j_h=\pm3/2$), so that $\mathcal{B}_{XH}= \left\{\left|+2\right\rangle,\left|+1\right\rangle,\left|-1\right\rangle,\left|-2\right\rangle\right\}$. The short-range interaction now takes the form:  
\begin{eqnarray}
& & H^{2D(SR)}_{hh}= \delta_{\bm K_\bot,\bm K'_\bot}\frac{3}{4}\Delta_0\frac{\left|\phi^{2D}_{h,1s}(0)\right|^2}{\left|\phi^{3D}_{1s}(0)\right|^2}I_{hh}
\left[ \begin{array}{cccc}
0&0&0&0\\
0&1&0&0\\
0&0&1&0\\	
0&0&0&0\\ \end{array} \right]  \nonumber\\
& & \hskip +1.5 cm =-\delta_{\bm K_\bot,\bm K'_\bot}\frac{2}{3}\Delta^{2D}_0 J_z S_z +cst.
\label{equation4}
\end{eqnarray} 
where the short-range 2D splitting is given by $\Delta^{2D}_0 = \frac{3}{4}\Delta_0\frac{\left|\phi^{2D}_{h,1s}(0)\right|^2}{\left|\phi^{3D}_{h,1s}(0)\right|^2}I_{hh}$.

Expressions (\ref{equation4}) show that: (\textit{i}) only excitons with the same center of mass wave-vector $\bm K_\bot$ can interact, (\textit{ii}) the short range exchange correction $\Delta^{2D}_0$ is independent of $\bm K_\bot$. In an infinite QW, the overlap integral is $I_{hh}=3/(2L_W)$ where $L_W$ is the QW width, so that : $ \Delta^{2D}_0 \approx \frac{9}{16}\Delta_0 \left(\frac{E^{2D}_B}{E^{3D}_B}\right)^2 \frac{a^{3D}_B}{L_W} $, showing that when the quantum well width decreases, the 2D short range exchange increases first, as is clear from figure \ref{FigureAI1} of Appendix I. This corresponds to the trend observed experimentally (see later, fig. \ref {Fig16}).  
To conclude on short-range exchange, in terms of effective Hamiltonian, describing the heavy-hole doublet by a pseudo-spin $\left(\left|3/2,\mp 3/2\right\rangle\equiv\left|1/2,\pm1/2\right.\left.\right\rangle_h\right)$ and using Pauli matrixes $\sigma_{e,i}$ and $\sigma_{h,i}$ ($i=x,y,z$) for electrons and holes, the most general form for type I QW of $D_{2d}$ symmetry is \cite{ivchenko}: 
\begin{eqnarray}
  H^{2D(SR)}_{hh}= \frac{\Delta_0}{2}\sigma_{e,z}\sigma_{h,z}+	\frac{\Delta_2}{4}\left(\sigma_{e,x}\sigma_{h,x}+\sigma_{e,y}\sigma_{h,y}\right)
	\label{equation5}
\end{eqnarray} 
where the small $\Delta_2$ term splits the $J=2$ heavy-hole exciton states. 
\begin{figure}[t]
	\centering
		\includegraphics*[width=.9\textwidth]{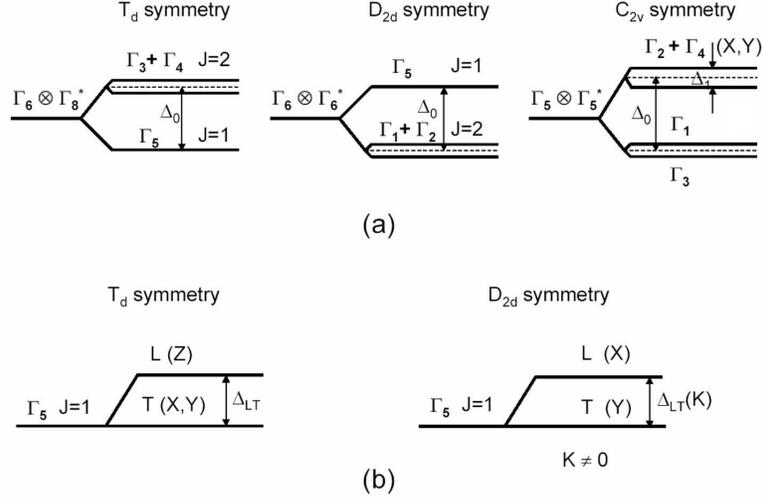}
		\caption{(a) Short range exchange splitting of exciton states in direct gap zinc-blende crystals ($T_d$ symmetry), and of heavy hole excitons in $D_{2d}$ or $C_{2v}$ $GaAs/AlGaAs$ [001] heterostructures. (b) Long range electron-hole exchange (non-analytic contribution) in crystals of $T_d$ symmetry, and $e_1-hh_1$ ground state heavy-hole exciton in $D_{2d}$ heterostructures (the splittings between the dark states - the $J=2$ excitons - can be neglected in most experimental situations). The notations of the representation $\Gamma_i$ are those of Koster tables \cite{Koster}.}
	\label{Figure1}
\end{figure}
In some QW structures, the confinement potential of electrons and holes does not occur in the same layer. For instance for GaAs/AlAs structures with sufficiently narrow GaAs layer, the holes are still confined in the GaAs layer, while the electrons are in the AlAs layer (the so-called type II quantum wells). The type II excitons (see Appendix I.2) are much more sensitive to the interface symmetry on which they are localized than type I excitons. Such interface has usually $C_{2v}$ symmetry. The short range exchange is thus modified accordingly. One obtains for $H^{2D(SR)}$ for type II QW \cite{ivchenko}: 
\begin{eqnarray}
 & & \hskip -1.5 cm H^{2D(SR)}_{hh}= \frac{\Delta_0}{2}\sigma_{e,z}\sigma_{h,z} \nonumber\\   
 & & +\frac{\Delta_1}{4}\left(\sigma_{e,x}\sigma_{h,y}+\sigma_{e,y}\sigma_{h,x}\right)  +\frac{\Delta_2}{4}\left(\sigma_{e,x}\sigma_{h,x}+\sigma_{e,y}\sigma_{h,y}\right) 
	\label{equation6}
\end{eqnarray} 
where an additional $\Delta_1$ term, often called \textit{anisotropic exchange}, couples the optically active ($J=1$) exciton states (belonging to $\Gamma_5$ representation of $D_{2d}$ in Koster notations \cite{Koster}), but not the $\Gamma_5$ to the $\Gamma_1$ nor to the $\Gamma_2$ states (of $D_{2d}$) ones. The optically active doublet (in $D_{2d}$) is thus split into two linear exciton states which dipoles are aligned along the [1,1,0] and the [1,-1,0] crystallographic directions. The figure 1(a) shows the evolution of the short range splitting when the symmetry of the semiconductor is lowered from $T_d$ to $D_{2d}$ and then to $C_{2v}$. 

\subsection{Long-range electron-hole exchange}
\label{sec:12}
In bulk material, it can be shown that, for $\bm K_G << \pi/a$ (a being the lattice parameter, and $\bm K_G$ the exciton center of mass wavevector), long-range exchange can be approximated by the operator with general matrix elements: 
\begin{eqnarray}
  -E_{\beta,\alpha}= \delta_{\bm K_G,\bm K'_G}\frac{4\pi e^2}{\epsilon_b}\left[ 3\frac{\left(\bm K_G\bm . \bm r_{\beta,\emptyset}\right) \left(\bm r_{\emptyset,\alpha}\bm . \bm K_G\right)}{K^2_G}-\bm r_{\beta,\emptyset}\bm . \bm r_{\emptyset,\alpha}\right]
	\label{equation7}
\end{eqnarray} 
where: $\left|\alpha\right\rangle \equiv \left|s,m_h,n,l,m\right\rangle $  and $\left|\beta\right\rangle \equiv \left|s',m'_h,n',l',m'\right\rangle $ are exciton states (\textit{n,l,m} indexes correspond to the 3D hydrogenic functions), $\left|\emptyset\right\rangle$ is the crystal ground state (without excitons);
\begin{small} $\bm r_{\alpha,\emptyset}\equiv\ \left\langle \alpha\right|\hat{\bm r}\left|\emptyset\right\rangle$
  $=\left(\phi^{3D}_{n,l,m}(r=0)\right)^*\left\langle s\left|\hat{\bm r}\hat{\bm K}\right|m_h\right\rangle$
 \end{small} is the dipole operator between the crystal ground state and the exciton state $\left|\alpha\right\rangle$, and $\hat{K}$ is the time reversal operator. Expressions (\ref{equation7}) show that : (\textit{i}) only excitons with the same center of mass wave-vector $\bm K_G$ can interact with long-range exchange. (\textit{ii}) the latter contains two contributions: the first one is the so-called non-analytic part, since its value changes depending on the way $\bm K_G$ goes to zero. It is responsible for the longitudinal-transverse splitting of excitons $\Delta_{LT}$. The second one is analytical, and is usually dropped. (\textit{iii}) the matrix elements $E_{\beta,\alpha}$ are non zero only between optically active states. As the angular momentum of valence states are $L=1$, only $\left|n,0,0\right\rangle$ states contribute to the non-analytic contribution to exchange. In cubic crystal, the energy of the longitudinal transverse splitting is given, for the ground-state $\Gamma_6\times\Gamma^*_8$ exciton by \cite{ivchenko} :
\begin{eqnarray}
 \Delta^{3D}_{LT}=\frac{16\pi e^2}{3\epsilon_b}\frac{\hbar^2 P^2}{E^2_g}\left|\phi^{3D}_{1s}(0)\right|^2 
	\label{equation8}
\end{eqnarray} 
where $P \equiv \left\langle iS \left|\hat{p}_z \right|Z \right\rangle /m_0$ is the usual Kane parameter. For GaAs $\Delta_{LT}\approx 0.1 meV$ \cite{Ulbrich}. 

In quantum wells structure, the calculation can be found in ref. \cite{mas}. For the lowest heavy-hole excitons, it leads, in the heavy-hole exciton basis $\mathcal{B}_{XH}$, to :  
\begin{eqnarray}
 H^{2D(LR)}_{hh}= \delta_{\bm K_\bot,\bm K'_\bot}\frac{1}{2}\Delta^{2D}_{LT}(\bm K_\bot)
\left[ \begin{array}{cccc}
0&0&0&0\\
0&1&-e^{-2i\varphi}&0\\
0&-e^{+2i\varphi}&1&0\\	
0&0&0&0\\ \end{array} \right] 
	\label{equation9}
\end{eqnarray} 
where  $\varphi=\varphi\left(\bm K_\bot\right)$ is the angle between $\bm K_\bot$ and the \textit{Ox} axis, 
\begin{eqnarray}
 \Delta^{2D}_{LT}(\bm K_\bot)=\frac{3}{8}\Delta^{3D}_{LT}  \frac{\left|\phi^{2D}_{h,1s}(0)\right|^2}{\left|\phi^{3D}_{1s}(0)\right|^2} K_\bot I_0(K_\bot)
	\label{equation10}
\end{eqnarray} 
and $I_0$ is a form factor, given by :\\ $I_0=\int^{+\infty}_{-\infty}\chi_{e,1}(z)\chi_{h,1}(z)dz\int^{+\infty}_{-\infty}\chi_{e,1}(z')\chi_{h,1}(z')dz'e^{-K_{\bot}\left|z-z'\right|}$ (the QW functions  $\chi_{e,\nu}(z)$ and $\chi_{h,\nu'}(z)$ are supposed as real here). For $K_\bot << \pi/L_W$, one can approximate $e^{-K_\bot\left|z-z'\right|}\approx 1$, and $I_0\approx \left|\left\langle \chi_{e,1}\left| \right.\chi_{h,1}\right\rangle\right|^2 $, \textit{i.e.} it reduces to the overlap of the electron and hole functions (in the infinite barrier model, $I_0=1$). Finally, we obtain the approximation:\begin{small}
$\Delta^{2D}_{LT}(K_{\bot})\approx \frac{3}{16} \Delta^{3D}_{LT} \left|\left\langle\chi_{e,1}\left|\chi_{h,1}\right.\right\rangle\right|^2\left(\frac{E^{2D}_B}{E^{3D}_B}\right)^2a^{3D}_B K_{\bot}$\end{small}. 
Contrary to the 3D case (see fig.1b), the 2D longitudinal transverse splitting is zero for $K_\bot=0$, and increases linearly with $K_\bot$. For instance, if $a^{3D}_B K_\bot\approx 0.1$, one can estimate  $\Delta^{2D}_{LT}\approx 40 \mu eV$ typically for GaAs/AlGaAs QWs of 2D character. Similarly to the short range exchange $\Delta^{2D}_0$, the long-range splitting $\Delta^{2D}_{LT}$ increases when the confinement increases (\textit{i.e.} when the well width decreases). As we shall see in section \ref{sec:34}, the long-range exchange interaction is at the origin of an important spin relaxation channel for excitons in type I quantum wells. 

\subsection{Exciton in magnetic field}
\label{sec:13}

It can be shown from the envelope function approach \cite{ivchenko} that the hamiltonian describing the exciton splitting in an external magnetic field $\bm B=(\bm B_\bot ,B_z)$ for a structure of $D_{2d}$ symmetry can be written as:
\begin{eqnarray}
& & \hskip +1.5 cm\mathcal{H}_B= \mathcal{H}_{e,B}+\mathcal{H}_{h,B} \nonumber\\
& & \mathcal{H}_{e,B}=\mu_B\left(g_{e,\parallel}B_z\hat{S}_z + g_{e,\bot}\bm B_\bot \bm . \hat{\bm S}_\bot \right) \nonumber\\
& & \mathcal{H}_{h,B}=\mu_B g_0 \left(\kappa \bm B \bm . \hat{\bm J}\right.+q \sum_{\alpha=x,y,z}\left.B_{\alpha} \hat{J}^3_{\alpha} \right) 
\label{equation11}
\end{eqnarray} 
where $\mu_B$ is the Bohr magneton, $g_0$ the free electron g-factor, and $\hat{\bm S}=(\hat{\bm S}_\bot,\hat{S}_z)$ and
$\hat{\bm J}=\left(\right.\hat{J}_x,\hat{J}_y,\hat{J}_z\left.\right)$ are the electron and hole spin and angular momentum operators respectively. The effective constants $\kappa$ and $q$ may differ from the one of free holes \cite{ivchenko}. Generally, one have $q << 1$. The electron g-factor is anisotropic ($g_{e,\bot}$, $g_{e,\parallel}$) due to the confinement along the growth axis \textit{Oz} of the QW which splits the heavy and light hole. 

In longitudinal magnetic field, for heavy-hole exciton, the matrix of $\mathcal{H}_B$, in the basis $\mathcal{B}'_{XH}$ $= \left\{\left|+1\right\rangle,\left|-1\right\rangle,\left|+2\right\rangle,\left|-2\right\rangle\right\}$, becomes : 
\begin{eqnarray}
	\mathcal{H}_{B,\parallel}= \frac{\hbar}{2}
\left[ \begin{array}{cccc}
\omega_-&0&0&0\\
0&-\omega_-&0&0\\
0&0&\omega_+&0\\	
0&0&0&-\omega_+\\ \end{array} \right]  
\label{equation12}
\end{eqnarray} 
with: $\hbar\omega_\pm=2\mu_B B_z \left[\frac{3}{2}\left(\kappa+\frac{9}{4}q\right)g_0\pm\frac{1}{2}g_{e,\parallel}\right]\equiv \mu_B B_z\left(-g_{h,\parallel}\pm g_{e,\parallel}\right)/2$, which allows us to define the exciton longitudinal g-factor $g^J_{exc,\parallel}$. This hamiltonian thus only splits the $J=1$ and $J=2$ heavy-hole exciton states. In order to analyse the exciton spectra under longitudinal magnetic field in magneto-optics experiments, one has to add the short-range exchange hamiltonian $H^{2D(SR)}_{hh}$, as will be shown later in section section \ref{sec:41}. 

In transverse magnetic field, the matrix of $\mathcal{H}_B$ takes now the form, in the same basis: 
\begin{eqnarray}
	\mathcal{H}_{B,\bot}= \frac{\hbar}{2}
\left[ \begin{array}{cccc}
0&0&\delta_{e,B}&\delta_{h,B}\\
0&0&\delta^*_{h,B}&\delta^*_{e,B}\\
\delta^*_{e,B}&\delta_{h,B}&0&0\\	
\delta^*_{h,B}&\delta_{e,B}&0&0\\ \end{array} \right]  
\label{equation13}
\end{eqnarray} 
with $\delta_{e,B}=g_{e,\bot}\mu_B\frac{B_x+iB_y}{\sqrt{2}}$ and $\delta_{h,B}=\frac{3}{2}g_0 q \mu_B \frac{B_x+iB_y}{\sqrt{2}}$ . The form (\ref{equation13}) with $q=0$ will be used together with  $H^{2D(SR)}_{hh}$ in section \ref{sec:42} to analyse exciton quantum beats, initially prepared in $\left|+1\right\rangle$ state by optical pumping.

\section{Optical orientation of exciton spin in Quantum Wells }
\label{sec:2}

Thanks to the development of stable ultrafast laser sources at the end of the 1980's, it has been possible to monitor directly in the time domain the carrier spin dynamics in semiconductors \cite{chao,vanderpoel2}.
Time-resolved polarization absorption measurements based on pump-probe techniques or time-resolved polarized photoluminescence (PL) experiments were extensively used to measure the spin relaxation of excitons in semiconductor quantum wells \cite{tackeuchi,freeman,damen}. 
These time-resolved techniques are very complementary tools to the well established measurements methods based on cw photoluminescence spectroscopy or Hanle type experiments \cite{optical,zerrouati}. 
The measurement of the circular polarization dynamics of the exciton luminescence after a circularly polarized ($\sigma^+$) pulsed laser excitation allows one to measure both (\textit{i}) the spin polarization of the exciton just after the $\delta$-like optical pump and compare it with the theoretical value given by the optical selection rules (see Appendix AI.3) and the band structure and (\textit{ii}) the decay time of the exciton spin polarization, which allows one to deduce the dominant spin relaxation mechanism. 

\begin{figure}[t]
	\centering
		\includegraphics*[width=.7\textwidth]{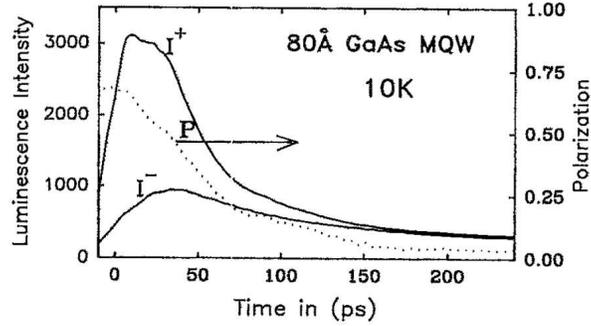}
		\caption{Intensities $I^+$ and $I^-$ and polarization $P= (I^+ - I^- )/(I^+ +I^- )$ of exciton luminescence as a function of time following picosecond excitation at the exciton energy. $I^+(I^-)$ corresponds to the intensity of $\sigma^+$ emission for $\sigma^+(\sigma^-)$ excitation \cite{damen}}
	\label{Fig1}
\end{figure}

\begin{figure}[t]
	\centering
		\includegraphics*[width=.7\textwidth]{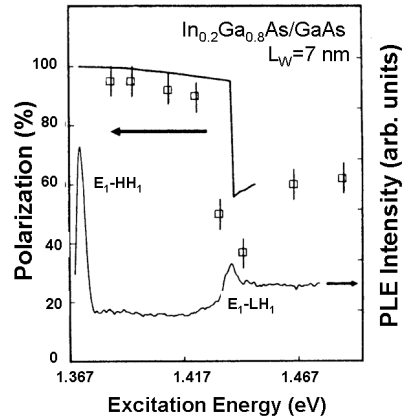}
		\caption{Photoluminescence Excitation (PLE) spectrum under cw stationnary excitation. The initial polarization degree $P_L(t=0)$ in dynamical experiments of the heavy-hole exciton luminescence as a function of the laser excitation energy is also displayed; ($\Box$): experimental data ; solid line: calculated values. $T=1.7 K$ \cite{dareys}.}
	\label{Fig2}
\end{figure}

Figure \ref{Fig1} displays for a $8 nm$ GaAs/AlGaAs Multiple Quantum Well (MQW) the time evolution of $I^+$ and $I^-$ for resonant excitation of the heavy-hole exciton XH ($I^+$ and $I^-$ correspond to the right circularly polarized ($\sigma^+$) luminescence component following a right ($\sigma^+$) or left ($\sigma^-$)    circularly polarized picosecond laser excitation respectively). Since only the heavy-hole exciton is excited (the  spectral width of the laser pulse is much smaller than the heavy-light hole splitting), the initial polarization $P_L(t=0)$ of the exciton luminescence is very large : $P_L(0)\approx70\%$ (the time-resolution of the streak camera used here as a detector is about $10 ps$). The circular polarization in figure \ref{Fig1} decays with a time constant of $\approx 50 ps$. The link between this decay time and the spin relaxation of exciton is not straightforward since several spin relaxation channels can occur simultaneously \cite{dareys,vinattieri}.
This will be discussed in detail in section \ref{sec:3}. 
Figure \ref{Fig2} presents the variation of the initial polarization $P_L(0)$ of the exciton luminescence as a function of the picosecond laser excitation energy in a compressively strained InGaAs/GaAs MQW ($L_W=7 nm$) \cite{dareys}.
When the incident photon energy is larger than the QW gap but smaller than the light-hole exciton transition (involving the E1 and $LH_1$ sub-bands), the initial polarization $P_L(0)$ is as high as 95\% (the time-resolution of the up-conversion time-resolved photoluminescence spectroscopy technique used here is about $1 ps$). This very high $P_L(0)$ value proves that the initial carrier thermalization process which occurs on a few hundreds of femtosecond time scale, leads to very minor carrier depolarization (at least for the conduction electrons under non resonant excitation). The calculated initial polarization for valence to conduction band transitions using the envelop function formalism and the effective Luttinger Hamiltonian is also plotted in figure \ref{Fig2} (full line) \cite{barrau}.
The variation of $P_L(0)$ versus the excitation energy is the result of valence band mixing. The mismatch around the ($E_1-LH_1$) excitation energy between the experiment and the calculation is just due to the fact that the latter does not take into account the absorption increase due to bound light-hole exciton state (XL) \cite{pfalz,oestreich}. As a fact, the XL oscillator strength may become stronger than the one of unbound $E_1-HH_1$ electron-hole pair states at the same energy (see Appendix I.3).
For a strictly resonant excitation of the light-hole exciton photoluminescence, the heavy hole exciton luminescence polarization can indeed be negative (opposite to the helicity of the excitation laser polarization), see the curve (3) in figure \ref{Fig3} for a $L_W=4 nm$ GaAs/AlGaAs multiple quantum well structure \cite{dareys2,pfalz}.
%

\section{Exciton spin dynamics in Quantum Wells }
\label{sec:3}
\begin{figure}[t]
	\centering
		\includegraphics*[width=.7\textwidth]{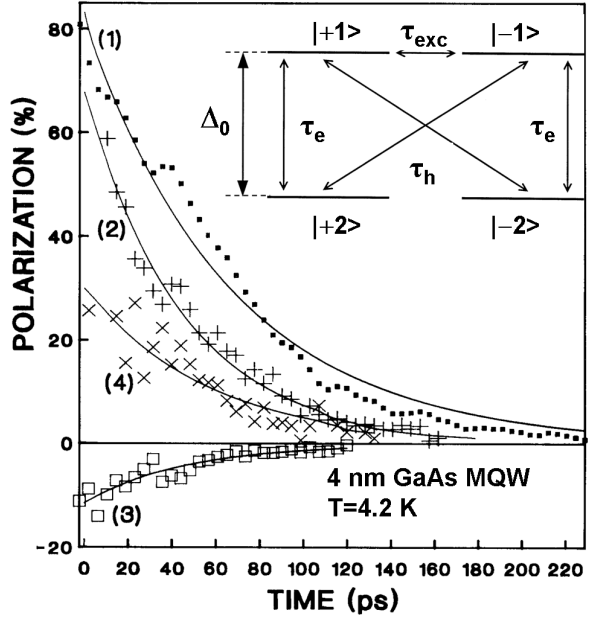}
		\caption{Circular polarization dynamics of the heavy-hole exciton XH  luminescence following a ($\sigma^+$)-polarized picosecond laser pulse. Four excitation energies : (1) $h\nu =XH+10 meV$ ; (2) $h\nu =XH+22 meV$ ; (3) $h\nu =XH+32 meV$, resonant with the light-hole exciton energy XL ; (4) $h\nu =XH+74 meV$  \cite{dareys2}.
Inset : Schematic diagram of the different exciton spin relaxation processes; $\tau_{exc}$, $\tau_{e}$ and $\tau_{h}$ represent the exciton, electron and hole spin relaxation time respectively (see text).}
	\label{Fig3}
\end{figure}
Exciton luminescence polarization studies in semiconductor QW have revealed the coexistence of two main mechanisms of exciton spin relaxation : the direct relaxation with simultaneous electron and hole spin flip due to the electron-hole exchange interaction \cite{mas} and an indirect one with sequential spin flips of the single particles (electron or hole), see the inset of figure \ref{Fig3}. The rate of exciton spin relaxation in this indirect channel is limited by the slower single particle spin-flip rate, which is typically the electron one \cite{andrada}. The relative efficiency of these mechanisms depends on the excitation conditions which can be \textit{resonant} (the energy of the polarized excitation photons is equal to the exciton energy) or \textit{non-resonant} (the photon excitation energy is typically above the QW gap energy $E_1-HH_1$). In the latter, the exciton spin dynamics is influenced by the exciton formation process \cite{amand}.

\subsection{Exciton formation in Quantum Wells }
\label{sec:31}

In bulk semiconductors, two exciton formation processes are usually considered: straight hot exciton photogeneration, with the simultaneous emission of an LO phonon, in which the constitutive electron-hole pair is geminate; or bimolecular exciton formation which consists of the random binding of electrons and holes under the Coulomb interaction.

The analysis of the initial polarization $P_L(t=0)$ of the exciton luminescence in time-resolved optical orientation experiments performed in GaAs/AlGaAs or InGaAs/GaAs QWs reveals precious information about this exciton formation process \cite{amand}. The idea is to measure the initial PL polarization $P_L(0)$ using an elliptically polarized laser beam, characterized by its degree of circular polarization defined as $P_E=(\Sigma^+-\Sigma^-)/(\Sigma^++\Sigma^-)$ 
  where $\Sigma^+$  and $\Sigma^-$  represents the intensities of the right and left circularly-polarized optical excitation components.
  \begin{figure}[t]
	\centering
		\includegraphics*[width=.7\textwidth]{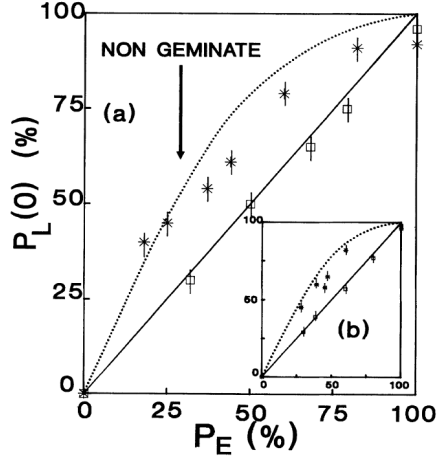}
		\caption{(a) Initial circular polarization degree of the exciton photoluminescence PL(t=0) versus the excitation light polarization degree $P_E$ of the picosecond laser pulse in a $L_W=7 nm$ $In_{0.2}Ga_{0.8}As$/GaAs multiple quantum well. The symbols represent the measured values for ($\Box$) quasi-resonant excitation : $h\nu =XH+4 meV$; ($\ast$) non-resonant excitation : $h\nu =XH+34 meV$. The continuous lines are, respectively, the calculated $P_L(0)$ values for a geminate and non-geminate (bimolecular) formation process.
(b) Similar analysis on a GaAs/$Al_{0.3}Ga_{0.7}As$ MQW ($L_W=4 nm$) ; ($\Box$): resonant excitation, $h\nu=XH$; ($\ast$): non-resonant excitation, $h\nu =XH+15 meV$ \cite{amand}}
	\label{Fig4}
\end{figure}
Figure \ref{Fig4} presents the experimental PL circular polarization degree $P_L(0)$ versus $P_E$ for non-resonant and resonant excitation conditions. The measurements are performed in a $L_W=7 nm$ InGaAs/GaAs multiple quantum well structure. The striking feature is that for non-resonant excitation ($h\nu =XH+34 meV <XL$, where XH is the heavy-hole exciton energy and XL the light-hole one), the initial PL polarization degree is higher than the excitation light polarization. In contrast, in resonant excitation ($h\nu\approx$  XH), below the QW gap, the behaviour is completely different: within the experimental accuracy the initial PL polarization is equal to the excitation light polarization, whatever the $P_E$  value is \cite{amand}.

In resonant excitation conditions, the excitons are formed from geminate pairs which keep their initial spin orientation; the initial PL polarization is thus : 
\begin{eqnarray}
	P_L(0)=P_E
	\label{equ1}
\end{eqnarray}
in agreement with the experimental results in figure \ref{Fig4}.

In non-resonant excitation conditions (above the QW gap), the polarized excitation pulse creates electron-hole pairs with a total spin $M=+1$ and $M=-1$. The proportions are $(1+P_E)/2$ and $(1-P_E)/2$ respectively. If the excitons are formed from spin-unrelaxed non-geminate pairs by a bimolecular formation process the initial excitonic populations on optically active and inactive spin states , $\left|\pm1\right\rangle$ and  $\left|\pm2\right\rangle$, are respectively :
$N_{\pm1}\propto(1\pm P_E)^2/4$ and $N_{\pm2}\propto(1- P_E^2)/4$ .

The coherence effects are neglected here since the electron and hole angular momenta are now uncorrelated. The initial polarization is then given by :
\begin{eqnarray}
P_L(0)=\frac{2 P_E}{1+P_E^2}\geq P_E
	\label{equ2}
\end{eqnarray}
an expression which shows that $P_L(0)$ is strictly higher than the polarization of the excitation light when $0<P_E<1$. This is due to the fact that the electron $\left|-1/2\right\rangle$  states are more populated than the $\left|+1/2\right\rangle$ ones, and have a higher probability to bind to a $\left|+3/2\right\rangle$  hole than a $\left|-3/2\right\rangle$ one. Expression (\ref{equ2}) is strictly independent both of the initially created electron-hole pair density and the value of the bimolecular formation coefficient \cite{robart,siantidis,deveaud,Piermarocchi}. Equations (\ref{equ1}) and (\ref{equ2})  are plotted in figure 1.4; the full and dotted lines correspond, respectively, to the geminate and non-geminate exciton formation process. The comparison of the calculated and experimental polarization leads to the conclusion that following non-resonant excitation most of the excitons are formed by the bimolecular process. 

\subsection{Exciton-bound hole spin relaxation}
\label{sec:32}

In contrast to bulk materials in which the hole spin relaxation time is very fast ($\leq 1 ps$, characteristic time of the wavector relaxation time) \cite{optical,lejeune}, the lifting of the degeneracy in $k=0$ between the heavy hole and light hole sub-bands in quantum wells yields a decrease of the valence band mixing and hence an increase of the hole spin relaxation time \cite{bastard, uenoyama, baylac, roussignol,damen2}. The exciton spin dynamics can thus be strongly affected by the hole single particle spin relaxation time, which occurs on the same time-scale as the direct exciton spin relaxation which connects the two optically active $\left|+1\right\rangle$  and $\left|-1\right\rangle$  exciton states (see section \ref{sec:34}) \cite{mas}. However the exciton-bound hole spin relaxation time is usually shorter than the free hole spin relaxation in QW since the exciton is composed of holes states with wavectors ranging up to $a^{2D}_B$ typically, and thus characterized by a significant valence band mixing.

Two experimental techniques have been used to measure directly the hole spin relaxation time  ($\tau_h$) within the 2D exciton \cite{amand,snoke}.

\subsubsection{a) Measurement of the hole spin relaxation time by monitoring the total luminescence intensity dynamics}
\label{sec:321}
This technique exploits the exciton bimocular formation process in the non-resonant excitation conditions. As shown in \ref{sec:31}, the exciton bimolecular formation process yields an initial population of the exciton in the optically inactive states $\left|\pm2\right\rangle$ with a proportion $(1- P_E^2)/2$.

\begin{figure}[t]
	\centering
		\includegraphics*[width=.7\textwidth]{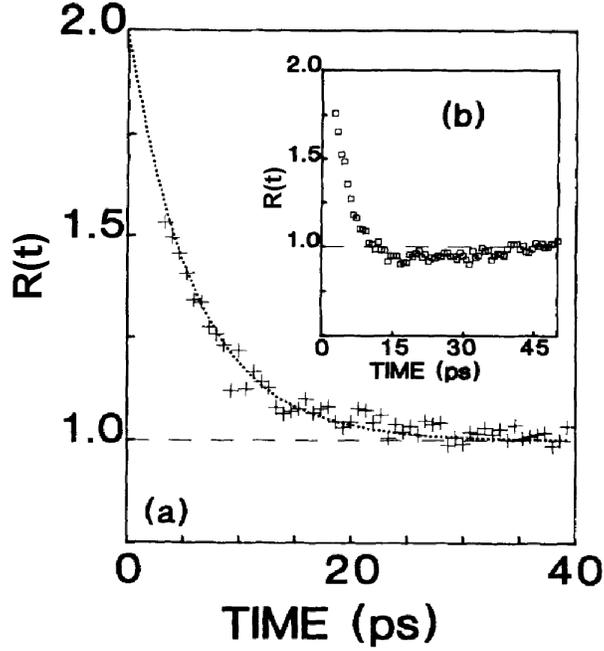}
		\caption{(a)	Time evolution of the ratio $R(t)=I_{\sigma^+}(t)/I_{\sigma^x}(t)$ where $I_{\sigma^+}(t)$ and $I_{\sigma^x}(t)$ are the total luminescence intensity following a ($\sigma^+$) circularly-polarized or ($\sigma^x$) linearly-polarized  excitation pulse in a $L_W=7 nm$ $In_{0.2}Ga_{0.8}As$/GaAs multiple quantum well. The solid line is an exponential fit of $R(t)$ according to $R(t)=2e^{-t/\tau_h}$, with $\tau_h=5.5 ps$.
(b)	Same measurements for a GaAs/$Al_{0.3}Ga_{0.7}As$ MQW ($L_W=4 nm$) \cite{baylac}. The experiment is performed at 1.7 K}
	\label{Fig5}
\end{figure}

Let us consider two different excitation conditions : first, a 100\% circularly  $\sigma^+$  light excitation; second a linearly polarized  $\sigma^x$ light excitation. In each case, the total luminescence intensity $I_{\sigma^+}$  and  $I_{\sigma^x}$ are recorded. These two measurement are performed at the same excitation energy (above the QW gap) and for the same excitation intensity. The ratio $R(t)=I_{\sigma^+}/I_{\sigma^x}$ is presented in figure \ref{Fig5} for the QW structures already presented in figure \ref{Fig4}. 
When the excitation is linearly polarized ($P_E=0$), the excitonic population is initially equi-distributed over the four states. Consequently, only half of the excitons are initially active and this does not change with time, since this distribution corresponds to the thermal equilibrium of the electronic excitations.

When the excitation is 100\% circular ($P_E=1$), only $\left|+1\right\rangle$ states are initially populated so that all the excitons are optically active at $t=0$. Consequently $R(0)=2$. The system will then tend to equalize the optically active and optically inactive excitonic population, due to the electron and hole single particle spin relaxation, so one expects a rapid decrease of $R(t)$ towards 1. The exciton spin-flip, governed by the exchange interaction between the electron and the hole (see section \ref{sec:34}), which changes the $\left|+1\right\rangle$ excitons into $\left|-1\right\rangle$ and vice-versa, is strictly inoperative in the time evolution of $R(t)$ which decays according to $R(t)=2e^{-t(1/\tau_h+1/\tau_e)}$. As the single particle exciton bound electron spin relaxation time is longer than the exciton bound hole spin relaxation time (as can be inferred later from section \ref{sec:33}), the evolution of $R(t)$ in figure \ref{Fig5} reflects directly the hole spin relaxation time. The fit of the experimental curves yields a hole spin relaxation time of $\tau_h= 5.5 ps$ and $\tau_h= 2.5 ps$ in the $InGaAs/GaAs$ and
$GaAs/AlGaAs$ QW structures presented. The advantage of this method is the direct measurement of $\tau_h$ in the exciton, independently of the determination of the exciton spin relaxation time. Moreover, it does not require the modelling of the exciton energy relaxation and the effective radiative recombination processes as they are identical for the two $I_{\sigma^+}$   and $I_{\sigma^x}$ recordings. As expected in resonant excitation conditions (geminate formation of excitons), $R(t)$ does not depend on time and equals 1 as expected \cite{amand}.

The energy dependence of the hole spin relaxation time has been studied by Baylac \textit{et al.}  with this technique \cite{baylac}. These authors found an hole spin relaxation time of  $\tau_h\approx15 ps$ for an excitation energy near the $InGaAs/GaAs$ QW gap $E_g$, dropping down to  $\tau_h\approx6ps$ for $h\nu >E_g+ 8 meV$ as a consequence of the valence band mixing and the increasing of the  electron-hole temperature with the increase of the excitation energy.

\subsubsection{b) Measurement of the hole spin relaxation with a two-photon excitation process }
\label{sec:322}

A different experiment allows direct measurement of the conversion rate of $J=2$ to $J=1$ excitons in GaAs Quantum Wells due to hole single particle spin relaxation \cite{snoke}. The experiment is basically as follows. First, $J=2$ excitons are created via two-photon infrared excitation, using an Optical Parametric Oscillator (OPO). Following the generation of the excitons, the single-photon recombination luminescence ($\approx$visible or near infrared) from the $J=1$ excitons is detected with a streak camera, which is completely insensitive in the infrared. Since the streak camera does not respond to the infrared exciting laser light, the $J=2$ excitons can be created by resonant excitation and observed immediately thereafter (after the conversion to $J=1$ states), without unwanted background from the laser light. This experiment relies on the fact that just as single photon emission from $J=2$ states is forbidden, two-photon absorption by J=1 excitons is forbidden but two-photon absorption by $J=2$ excitons is allowed.
\begin{figure}[t]
	\centering
		\includegraphics*[width=.7\textwidth]{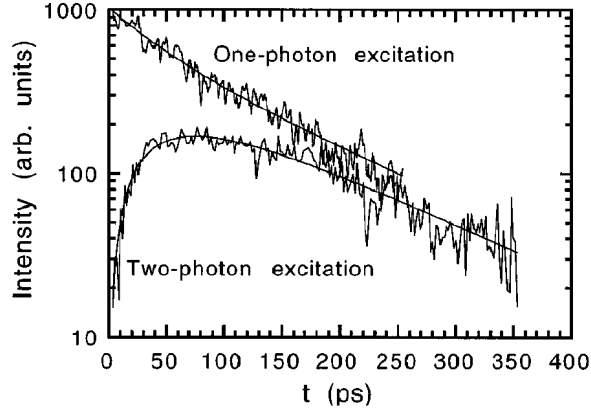}
		\caption{Lower curve : Luminescence intensity dynamics of the heavy-hole exciton $J=1$ in a $L_W=3 nm$ GaAs/AlGaAs QW, following the generation in the $J=2$ spin state by a $100 fs$ circularly polarized laser pulse at $1471 nm$.
Upper curve : Luminescence intensity dynamics of the heavy-hole exciton $J=1$ in the same QW, following the generation in the $J=1$ spin state by a $730 nm$ laser pulse (the relative intensity scales of the two curves are arbitrary) \cite{snoke}.}
	\label{Fig6}
\end{figure}
The lower curve of Fig. \ref{Fig6} shows as a function of time the XH (J=1) exciton luminescence at $730 nm$, from a $3nm$ quantum well at $2 K$, excited by circularly polarized OPO light \textit{i.e.} following two-photon excitation of the 1$s$ heavy-hole resonance. The rise time of the luminescence intensity after the two-photon excitation is mainly governed by the hole spin relaxation time  $\tau_h$ (which is shorter than the single particle electron spin relaxation time \cite{andrada}, see \ref{sec:33}). On the basis of simple rate equations for the $J=1$ and $J=2$ exciton states, Snoke \textit{et al.} concluded that the time scale for the hole spin-flip process in a narrow ($L_W=3 nm$) GaAs QW is of the order of $60 ps$ \cite{snoke}, which corresponds here to resonantly created XH excitons.

\subsection{Exciton-bound electron spin relaxation}
\label{sec:33}
The exciton-bound electron spin relaxation has been calculated by E. A. de Andrada e Silva and G. C. La Rocca taking into account the conduction band splitting due to the spin orbit interaction \cite{andrada}. They have shown that the off-diagonal matrix element between optical active and inactive exciton states that differ only with regard to the electron spin direction can be represented by an effective  magnetic field that changes randomly as the exciton is elastically scattered and relaxes its spin. The exchange splitting  $\Delta_0$ between the optical active and inactive states acts as a constant external magnetic field, reducing the electron spin relaxation rate. The estimated rate of the bound electron spin flip agrees well with values obtained from fitting the experimental data (see \ref{sec:34}) \cite{vinattieri,dareys}.
\begin{figure}[t]
	\centering
		\includegraphics*[width=.7\textwidth]{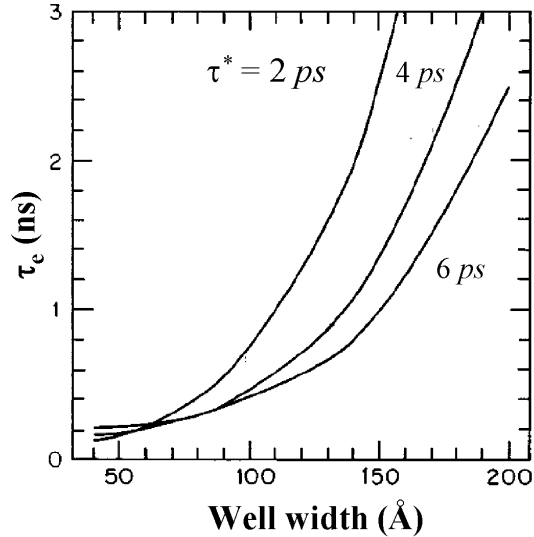}
		\caption{Well width variation of the exciton-bound electron spin relaxation time $\tau_e$ for different values of the elastic momentum scattering time $\tau^*$ in $GaAs/AlGaAs$ quantum wells \cite{andrada}.}
	\label{Fig7}
\end{figure}

This spin relaxation rate $W_e=1/2 \tau_e$  writes \cite{andrada} :
\begin{eqnarray}
	W_e=\frac{4\alpha^2K^2}{\hbar}\frac{\tau^*}{1+(\Delta_0 \tau^*/\hbar)^2}
	\label{equ3}
\end{eqnarray}
where   $\tau^*$ is the exciton elastic momentum scattering time,  $\Delta_0$ is the exchange splitting between the optical active and inactive exciton states, $K$ is the exciton wave vector and  $\alpha$ a constant depending on spin-orbit interaction in the conduction band.

This means that the exciton-bound electron spin dynamics presents a motional narrowing type of relaxation analogous to the D'Yakonov-Perel free-electron spin relaxation \cite{dp,dyakonov2}. Figure (\ref{Fig7}) displays the well-width dependence of this exciton-bound electron spin relaxation $\tau_e$ for different values of the elastic momentum scattering $\tau^*$ in $GaAs/AlGaAs$ QW. The spin relaxation time increases with the well width due to the corresponding decrease in the average spin-orbit splitting in the conduction band that the bound electron feels.

Except in the narrow well limit, we observe the usual motional narrowing behaviour with the exciton-bound spin-relaxation time roughly inversely proportional to the momentum scattering time. 
Experimental investigations of the exciton-spin dynamics in high-quality $GaAs/Al_xGa_{1-x}As$ multiple quantum wells ($x=0.3$ and $L_W=15 nm$) have determined through detailed fitting procedures that the exciton-bound electron-spin relaxation rate lays in the range $3\times10^8 s^{-1}<W_e< 3\times10^9 s^{-1}$ \cite{vinattieri,dareys}, in agreement with the calculated values plotted in figure \ref{Fig7}.

\subsection{Exciton spin relaxation mechanism}
\label{sec:34}

\subsubsection{a) The MAS mechanism}\label{sec:341}

\begin{figure}[t]
	\centering
		\includegraphics*[width=.7\textwidth]{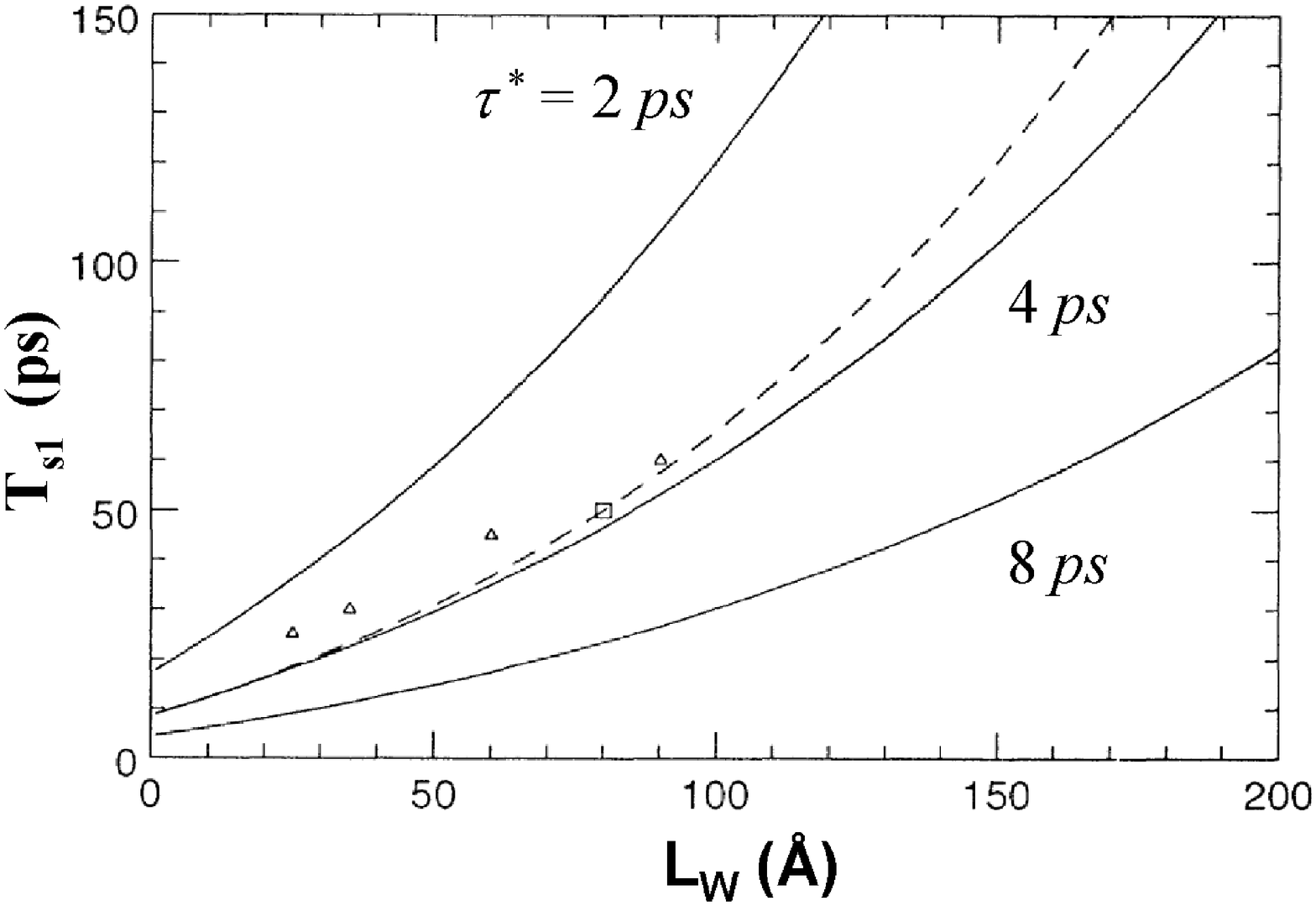}
		\caption{Calculated exciton spin relaxation time ($T_{s1}\equiv \tau_{exc}$) versus well width for different values of the momentum scattering time in $GaAs/AlGaAs$ quantum wells \cite{mas}. The experimental points are from \cite{roussignol} ($\triangle$) and \cite{damen} ($\Box$).}
	\label{Fig8}
\end{figure}

The main exciton spin depolarization mechanism in QW occurs via the exchange Coulomb interaction between the electron and the hole. The theory of this mechanism has been developed by M. Z. Maille, E. A. de Andrada e Silva and L. J. Sham (MAS process) \cite{mas}. The restriction of the heavy-hole exciton long-range exchange (equation \ref{equation9}) to the $J=1$ doublet ($\left|\pm 3/2,\mp 1/2\right\rangle\equiv\left|\pm 1/2\right\rangle_{exc}$) allows us to describe its spin dynamics using the pseudo-spin formalism. From equation \ref{equation9}, the restriction of $H^{2D(LR)}_{hh}$ takes the form: $\tilde{H}^{2D(LR)}_{hh}=\Delta^{2D}_{LT}/2+\bm \Omega_{LT} (\bm K_\bot)\bm . \bm \hat{S}_{ex}$, where $\bm \hat{S}_{ex}=(\hbar/2)(\sigma_x,\sigma_y,\sigma_z)$ is the XH exciton pseudo-spin, and $\bm \Omega_{LT}(\bm K_\bot)=-\Delta^{2D}_{LT}/(2\hbar)(cos2\varphi,sin2\varphi,0)$ is a precession vector. 
Letting $\bm S_{ex}= <\hat{\bm S}_{ex}>$ be the average exciton pseudo-spin, its time evolution is given by: $d\bm S_{ex}/dt= \bm \Omega_{LT} (\bm K_\bot)\times \bm S_{ex}$. 
The process may then be viewed as due to exciton spin precession in a fluctuating effective-magnetic field located in the well interface plane. The magnitude and direction of this field depends on the centre of mass momentum $\bm K_\bot$, and vanishes for $K_\bot=0$ states. Its correlation time corresponds to the exciton momentum scattering time $\tau^*$. The scattering of the centre of mass momentum creates a random effective magnetic field, responsible for the exciton spin relaxation, in the same manner as any other motional narrowing spin-flip processes, with the characteristic dependence of the spin-relaxation time on $(\tau^*)^{-1}$. The inverse exciton spin relaxation time ($\tau_{exc}$, often labeled $T_{s1}$) is given, provided that $\Omega_{LT}(K) \tau^*<<1$ holds, by:
\begin{eqnarray}
	\frac{1}{T_{s1}}\approx \left\langle \Omega_{LT}^2\right\rangle \tau^*
	\end{eqnarray}
where the square of the precession angular frequency $\Omega^2_{LT}(K)$ is now averaged on the whole exciton population. 
The time $T_{s1}$ is called \textit{longitudinal spin relaxation time} ; it corresponds to the relaxation between the $\left|+1\right\rangle$ and $\left|-1\right\rangle$ exciton states (\textit{i.e} circular depolarization time of  exciton luminescence). MAS have also calculated the \textit{transverse spin relaxation time} $T_{s2}$ which corresponds to the relaxation time of the coherence between $\left|+1\right\rangle$ and $\left|-1\right\rangle$ states. In the motional narrowing regime, and at low exciton density (see later, \ref{sec:62}), $T_{s2}\approx 2T_{s1}$. The transverse exciton spin relaxation can be measured by recording the decay time of the exciton linear depolarization in optical alignment experiments (see section \ref{sec:42}) \cite{mariePRL}.
Figure 	\ref{Fig8} presents the calculated exciton spin relaxation time $T_{s1}$ as a function of the well width for different values of the exciton momentum scattering time  $\tau^*$.  The quantum well confinement enhances the exchange interaction over its value in bulk, as shown in section \ref{sec:12}. The long-range exchange interaction is found to be the dominant contribution to the spin-relaxation process, whereas the short-range contribution is rendered less important by the need of assistance of the heavy- and light-hole coupling in the valence band that is reduced by the sub-band formation in lower-dimensional system \cite{mas}.
%

\subsubsection{b) Measurement of the MAS spin relaxation time }\label{sec:342}

The measurement of the exciton spin relaxation time  $\tau_{exc}$ (or $T_{s1}$) requires a fitting procedure of the experimental data, taking into account the single particle spin relaxation time of electrons ($\tau_e$)  and holes ($\tau_h$) within the exciton and the direct exciton spin relaxation time ($\tau_{exc}$), see the inset in figure 	\ref{Fig3}. If the experiments are performed in non-resonant excitation conditions, the model must also take into account the bimolecular formation process (see \ref{sec:31}).

The rate equations describing the different exciton spin states $\left|M\right\rangle$ populations $N_M$, (where $M=\pm1$, $\pm2$) once they have been created, are written, in the following equation, as a function of the electron, hole and exciton spin transition rates, $W_e=1/2 \tau_e$, $W_h=1/2 \tau_h$ and $W_{exc}=1/2 \tau_{exc}$ respectively, the latter being driven by the exchange interaction \cite{mas,vinattieri,dareys} :
\begin{eqnarray}
& & \hskip +2 cm \frac{d}{dt}
	\begin{pmatrix}	N_2\\	N_1\\	N_{-1}\\N_{-2}\end{pmatrix}
	=\bm{\left[W\right]}\begin{pmatrix}	N_2\\	N_1\\	N_{-1}\\N_{-2}\end{pmatrix} \nonumber\\
& & \hskip -1 cm \begin{small} \bm{\left[W\right]}=\begin{pmatrix}
	-W_{eh}&W_e&W_h&0\\
	W_e&-(1/\tau_r+W_{exc}+W_{eh})&W_{exc}&W_h\\
	W_h&W_{exc}&-(1/\tau_r+W_{exc}+W_{eh})&W_e\\
	0&W_h&W_e&-W_{eh}
	\end{pmatrix} 
	\end{small}
\label{equ4}	
\end{eqnarray}
where $W_{eh}\equiv W_e+W_h$ and $\tau_r$ is the recombination time.
\begin{figure}[t]
	\centering
		\includegraphics*[width=.7\textwidth]{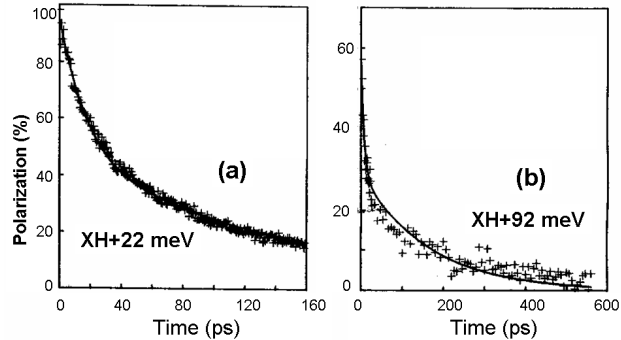}
		\caption{Time evolution of the experimental luminescence circular polarization for two excitation energies in a InGaAs/GaAs QW structure :  (a) $h\nu =XH+22 meV$ and (b)  $h \nu=XH+92 meV$ , \textit{i.e.} larger than the light-hole exciton energy. The solid line corresponds to the fit with the model presented in the text \cite{dareys}.}
	\label{Fig9}
\end{figure}
The calculated polarization is simply given by $P_{cal}(t)=\frac{N_1-N_{-1}}{N_1+N_{-1}}$. The straight lines in figures \ref{Fig9}.a and \ref{Fig9}.b correspond to least square fits of the experimental curves of the exciton PL circular polarization dynamics assuming the previous rate equations for a $L_W=7 nm$ InGaAs/GaAs QW structure. The depolarization dynamics are well described by an exciton spin relaxation ($ \tau_{exc}\approx 58ps$ and $79 ps$ respectively for the two excitation conditions), and a shorter time ($17ps$ and $7 ps$ respectively) which is identified as $\tau_h$. The fit gives a third time ($\tau_e$) much longer, greater than $1 ns$ : as a matter of fact, the fit is not very sensitive to this third time. It is impossible to fit the data with only  $\tau_e$ and  $\tau_h$: a finite excitonic spin relaxation  $\tau_{exc}$ is compulsory to get a good agreement. But the excitonic spin relaxation time alone can not explain the polarization decay as it leads to a calculated curve which is mono-exponential whereas the experimental ones are not.
\begin{figure}[t]
	\centering
		\includegraphics*[width=.7\textwidth]{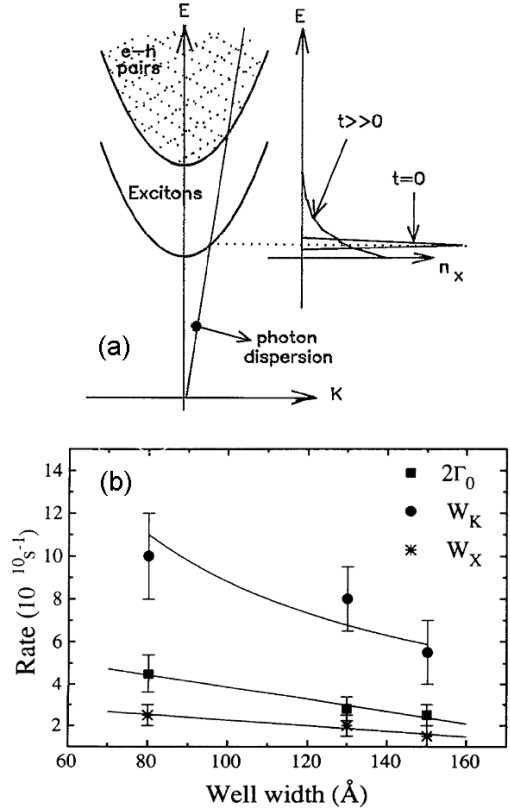}
		\caption{(a) Energy band diagram in a two-particle picture, showing the initial created exciton distribution and a thermalized exciton distribution \cite{damen}. 
		(b) Radiative recombination rate ($2 \Gamma_0$)   of the exciton population at $K_\parallel=0$, exciton effective scattering rate with phonons ($W_K$) and exciton spin relaxation time ($W_x=1/2 \tau_{exc}$). These rates are obtained from fits to the measured exciton polarization PL dynamics at $T=12 K$ in GaAs/AlGaAs QW structures\cite{vinattieri}}
	\label{Fig10}
\end{figure}
In contrast to what could be expected the modeling of the exciton spin depolarization dynamics measured by luminescence spectroscopy in strictly resonant excitation is not straightforward \cite{vinattieri,damen}. The measured temporal dynamics of resonantly-excited luminescence is determined by the relaxation, thermalization and recombination dynamics of this initial non-thermal distribution of exciton. Figure \ref{Fig10}.(a) schematically displays the relevant energy diagram in the two particle or exciton representation \cite{damen}.

The absorption of photons takes place only within the homogeneous width of the exciton. The homogeneous exciton linewidth of high quality MQW samples is usually less than $kT$ (even at $10 K$). As the photoexcited cold excitons thermalize, their distribution becomes wider than the initial distribution so that fewer excitons remain within the homogeneous linewidth of the exciton with increasing time. Since only excitons within the homogeneous linewidth couple to light as a consequence of the wave vector conservation, this leads to a decrease in the luminescence intensity even though the total number of excitons has not decreased \cite{deveaud2,vinattieri}. This process has thus to be taken into account in addition to the spin relaxation mechanisms previously described. Vinattieri and co-workers performed a comprehensive investigation of the dynamics of resonantly excited excitons in $GaAs/AlGaAs$ QW on picosecond time-scales \cite{vinattieri}. With systematic multi-parameter fits, they managed to extract the different relaxation rates, see figure \ref{Fig10}.b. They found  $W_x=1.5\times 10^{10} s^{-1}$, $W_h=0.7\times10^{10} s^{-1}$, and $3 \times10^8s^{-1} < We < 3\times 10^9 s^{-1}$ for a $15 nm$ $GaAs/AlGaAs$ QW structure.

Non-degenerate, spectrally, and spin-resolved differential transmission experiments allow also the determination of the different spin-relaxation times within the exciton \cite{ostatnicky,soleimani}. In these pump-probe experiments, the picosecond  $\sigma^+$   pump pulse is resonant with the $\left|+1\right\rangle$  QW excitons formed with $+3/2$ heavy holes $(hh)$ and $-1/2$ electrons ; the non-degenerate probe pulse measures the absorption at the light hole ($lh$) transition. The transmission change of this probe pulse as a function of time with polarization  $\sigma^-$   is not sensitive to the population at the $(hh)$ states with the angular momentum $+3/2$ but it is sensitive to the population of electrons with spins $-1/2$. In the same way, the  $\sigma^-$  probe  transmission change  at the hh excitonic transition is only sensitive to the population of $+1/2$ electrons and $-3/2$ holes. The last two bands are not initially populated by the pump pulse and, therefore, the population of these states results from electron or hole spin-flip processes. Using this experimental technique,  it is possible to extract unambiguously the time constants corresponding to the spin relaxation of one of the three types of quasiparticles. Ostatnicky et al measured for instance $\tau_e=250 ps$ and  $\tau_h=30 ps$ in a $10 nm$ thick GaAs MQW structure \cite{ostatnicky}. 
\begin{figure}[t]
	\centering
		\includegraphics*[width=.7\textwidth,height=.9\textwidth]{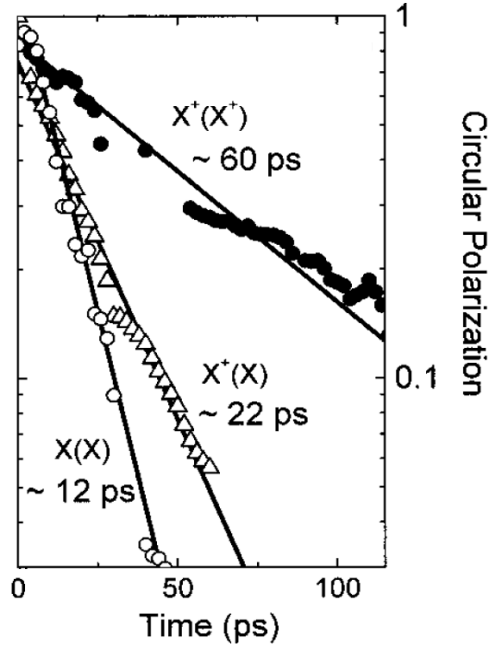}
		\caption{Circular photoluminescence polarization dynamics in a modulation $p$-doped CdTe/CdMgZnTe QW ($L_W=7.7 nm$), $T=10 K$. The neutral exciton $X$ PL dynamics is recorded after a resonant excitation of $X$ [noted $X(X)$], the $X^+$ PL dynamics is recorded after a resonant excitation of $X$ [noted $X^+ (X)$], and the $X^+$ PL dynamics after a resonant excitation of $X^+$ is also studied [noted $X^+(X^+)$] \cite{vanelle}.}
	\label{Fig11}
\end{figure}

The spin dynamics of neutral ($X$) and positively charged excitons ($X^+$ made of a hole singlet and one electron) have been measured and compared in modulation $p$-doped CdTe/CdMgZnTe quantum wells \cite{vanelle}. Thanks to the larger binding energy of the charged exciton ($X^+$) in II-VI QWs compared to the one in GaAs QWs  \cite{kheng,chen,dzhioev}, it is possible to study the neutral exciton $X$  PL dynamics after a resonant excitation of $X$ [noted $X(X)$ in the following],  the $X^+$ PL dynamics after a resonant excitation of $X$ [noted $X^+(X)$], and the $X^+$  PL dynamics after a resonant excitation of $X^+$ [noted $X^+(X^+)$]. Figure \ref{Fig11} illustrates the corresponding decay of the PL circular polarization for the three configurations. The neutral excitonic polarization [$X(X)$ spectrum] decreases with a time constant of $12 ps$, four times shorter than in typical III-V QW's of comparable sizes because of the larger exchange interaction (see section \ref{sec:1}) \cite{vinattieri}. As the neural excitons $X$ are created resonantly, \textit{i.e}., without kinetic energy, this time reflects mainly the excitonic spin-flip time $\tau_{exc}$ \textit{i.e.}, the simultaneous spin flip of the electron and the hole within the neutral exciton \cite{mas}. The $X^+(X^+)$ circular polarization decreases with a significantly longer time ($\approx60 ps$). As the $X^+$ is formed with two heavy holes of opposite spin (\textit{i.e.}, $m_h=+3/2$ and $m_h=-3/2$ respectively), the electron-hole exchange cancels in this charged exciton complex, so that the polarization of the charged excitons reflects the spin relaxation  $\tau_{e}$ of the electron only. The polarization decay time of the $X^+$ generated via X states [$X^+(X)$ spectrum] is intermediate, with an average time constant $\approx22 ps$. This intermediate behavior originates directly from the continuous creation of $X^+$ by the neutral $X$ : the $X^+$ created at short times $t<\tau_{exc}$ result from highly polarized $X$ and those retain their polarization for quite a long time ($\tau_e$), while nonpolarized $X^+$ are generated at slightly longer delays from excitons that have already lost their spin orientation. The fact that the $X^+(X)$ exhibits a strong initial polarization shows that the creation of $X^+$ via $X$ states does not affect the spin orientation.

\subsubsection{c) Electric field dependence of the exciton spin relaxation time }
\label{sec:343}
\begin{figure}[t]
	\centering
		\includegraphics*[width=.7\textwidth]{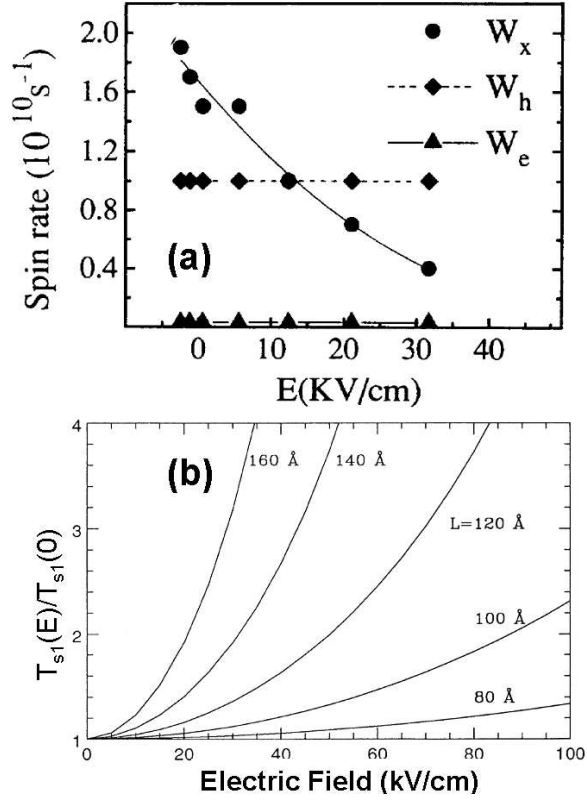}
		\caption{(a)	Measured dependence of the different spin relaxation rates $W_x$, $W_h$, and $W_e$ on the applied electric field for a $15 nm$ GaAs/AlGaAs QW, $T=20 K$ \cite{vinattieri}.
(b)	Calculated dependence of the exciton-spin relaxation time for various well widths \cite{mas}.}
	\label{Fig12}
\end{figure}
An electric field applied along the QW growth axis will increase the separation between the electron and the hole within the exciton. The reduction of the overlap between the electron and the hole wavefunction will yield a decrease of the long-range part of the exchange interaction (see section \ref{sec:12}). As a result, the exciton spin relaxation rate decreases when the applied electric field increases (fig \ref{Fig12}.a ). The measured variation of $W_x=1/2\tau_{exc}$ is in rather good agreement with the calculated one (fig \ref{Fig12}.b) \cite{vinattieri,mas}.

\subsubsection{d) Magnetic field dependence of the exciton spin relaxation time }
\label{sec:344}

Applying a moderate magnetic field along the QW growth direction also inhibits the exchange-driven exciton spin relaxation in GaAs QW because of the magnetic field induced Zeeman splitting $\Omega_0$ of the optically active exciton states \cite{tsitsishvili}. The magnetic field dependence of the longitudinal exciton spin relaxation $T_{s1}$ writes \cite{mas} :
\begin{eqnarray}
	\frac{1}{T_{s1}}=\left\langle \Omega_{LT}^2 \right\rangle \frac{\tau^*}{1+(\Omega_0\tau^*)^2}
	\label{equ5}
\end{eqnarray}
 where $\tau^*$ is the exciton wave vector relaxation time.
\begin{figure}[t]
	\centering
		\includegraphics*[width=.7\textwidth,height=.9\textwidth]{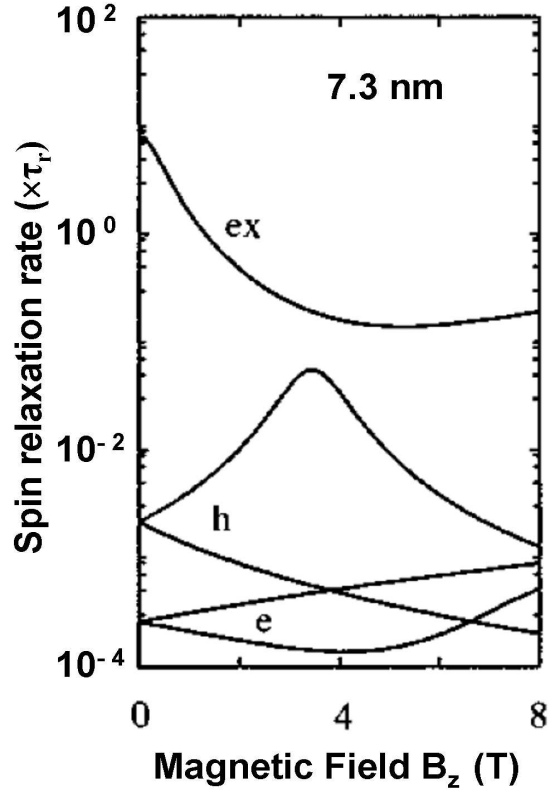}
		\caption{Measured spin relaxation rates of exciton (ex), holes (h) and electrons (e) in a longitudinal magnetic field $B_z$ for a $L_W=7.3 nm$ GaAs/AlGaAs QW structure. The rates are normalized to the total exciton population decay rate $\tau^{-1}_r$ \cite{harley}.}
	\label{Fig13}
\end{figure}
The magnetic field dependence of spin relaxation of heavy-hole exciton has been measured by Harley \textit{et al.} using cw magneto-photoluminescence experiments \cite{harley}. Figure \ref{Fig13} displays the spin relaxation rate of exciton (exc), electron (e) and hole (h) deduced from least square fitting of the experimental data (using equations 	(\ref{equ4})) for a $L_W=7.3 nm$ GaAs/AlGaAs MQW \cite{dareys,vinattieri}. The strong reduction of the exciton spin relaxation with applied magnetic field is clearly observed. These results have been confirmed by direct time-resolved measurements of the exciton spin dynamics using a dynamical Kerr Rotation experiment \cite{worsley}.

\section{Exciton exchange energy and g-factor in quantum wells }
\label{sec:4}

The exciton exchange energy and $g$ factor are strongly modified compared to bulk values because of the confinement of the electron and hole wavefunctions along the QW growth direction. Both cw and time-resolved optical spectroscopy techniques have been used to measure these parameters in various QW structures \cite{harley,puls,amand2,dyakonov}.

\subsection{Exchange interaction of excitons and g-factor measured with cw photoluminescence spectroscopy }
\label{sec:41}

\subsubsection{a) Exchange energy}
The value of the short-range exchange interaction in GaAs QW was first deduced from the measurements of the degree of circular polarization versus magnetic field of photoexcited luminescence \cite{blackwood}. The results presented below show evidence of exciton level crossings, which have been analysed to give the short-range exciton exchange energy, which is about $\Delta_0\approx 150 \mu eV$ for narrow GaAs/AlGaAs QW ($L_W \lesssim 5 nm$).
\begin{figure}[t]
	\centering
		\includegraphics*[width=.7\textwidth]{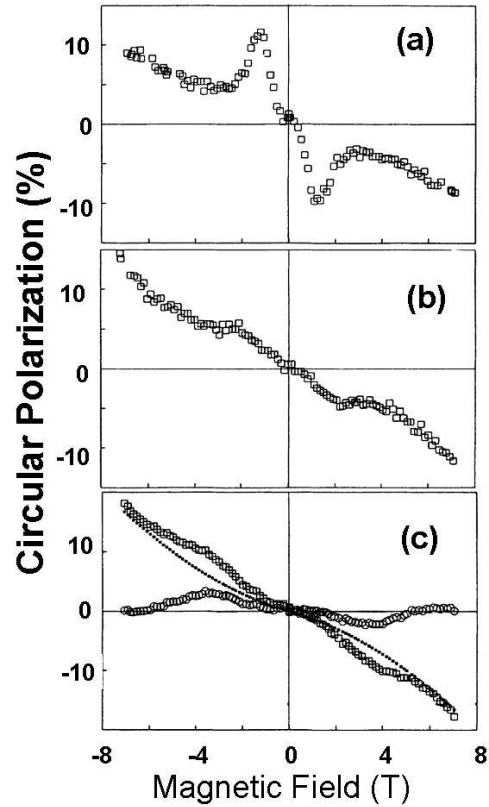}
		\caption{Circular polarized cw photoluminescence for GaAs/AlGaAs multiple quantum well samples with well widths (a) $2.5 nm$, (b) $5.6 nm$ and (c) $7.3 nm$ \cite{blackwood}}
	\label{Fig14}
\end{figure}
The elegant technique used by Blackwood \textit{et al.} relies on the measurement of the degree of circular polarization of the luminescence as a function of the applied magnetic field $B_z$ (applied along the growth axis), under non-resonant linearly-polarized cw laser excitation. Figure \ref{Fig14} presents the variation of the circular polarization degrees $P$ as a function of $B_z$ for three QW structures with different well widths \cite{blackwood}. There is a general monotonic increase of $\left|P\right|$ with applied field, the sign depending on the direction of the field, with a superimposed peak at a field which varies with QW width. This peak is due to magnetic field induced exciton level crossing (see fig. \ref{Fig15}). As the excitation is non-resonant (photogeneration of electron-hole pairs in the QW continuum), the bimolecular formation process of exciton will yield heavy-hole excitons in each of the four spin states 
($\left|M\right.\left.\right\rangle\ = \left|+1\right.\left.\right\rangle,\left|-1\right.\left.\right\rangle,\left|+2\right.\left.\right\rangle,\left|-2\right.\left.\right\rangle$) with equal probability and the relative populations of the states under cw excitation will be determined by the balance of recombination processes and phonon-assisted relaxation between the levels. Thus populations of the two optically allowed levels will tend towards the Boltzmann thermal distribution, with the degree of thermalization depending on the relaxation rates between the levels. If these rates vary smoothly the population difference of the optically allowed levels will increase steadily with applied field. However, the transition rate between a pair of levels will increase sharply if their energies become equal, because the transition can then occur without the intervention of a phonon. This will be reflected in an anomaly (presence of a peak) in the population difference of the optically allowed levels and therefore in the degree of circular polarization P of the integrated luminescence. Referring to figure \ref{Fig15}, there are in general two fields at which levels cross.
\begin{figure}[t]
	\centering
		\includegraphics*[width=.7\textwidth]{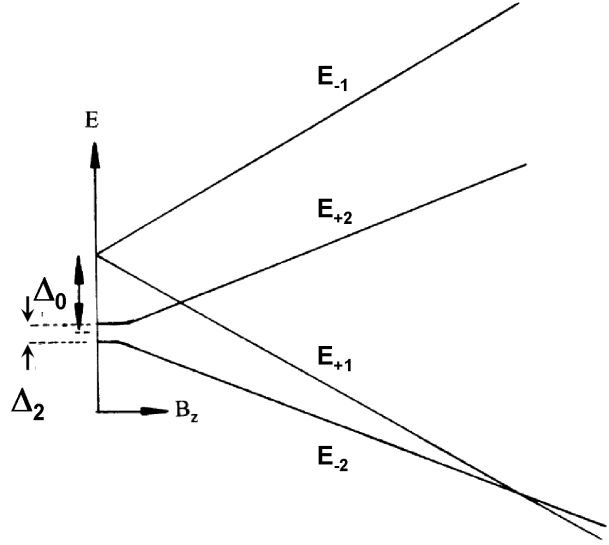}
		\caption{Exciton energy levels as a function of the longitudinal magnetic field $B_z$. $E_{\pm 1}$ and $E_{\pm 2}$) correspond to the $\left|\pm1\right\rangle$ optically active exciton states and $\left|\pm2\right\rangle$ non-optically active states respectively \cite{blackwood}.}
	\label{Fig15}
\end{figure}
The calculation of the position of these level crossing allows one to estimate the zero-field exchange energy $\Delta_0$.   The effective Hamiltonian representing the interaction of a 1$s$ exciton with a longitudinal magnetic field $B_z$ can be written generally, according to (\ref{equation6}, \ref{equation12}), as:
\begin{eqnarray}
	& & \hskip -0.7 cm H_{ex}=H^{2D(SR)}_{hh}+\mathcal{H}_{B\parallel}   \nonumber\\
	& & = 2\Delta_0 S_{e,z}S_{h,z}+ \Delta_1(S_{e,x}S_{h,y}+S_{e,y}S_{h,x})+ \Delta_2(S_{e,x}S_{h,x}+S_{e,y}S_{h,y}) \nonumber\\ 
	& & \hskip +4 cm +\mu_BB_z(g_{e,\parallel}S_{e,z}+g_{h,\parallel}S_{h,z})   \label{equ6}    
\end{eqnarray}
where $\bm S_{e}$ is the electron spin operator and $\bm S_{h}$ is an effective spin operator representing the two heavy-hole states $\left|\pm 3/2\right.\left.\right\rangle \equiv \left|\mp1/2\right.\left.\right\rangle_h$. The parameters $g_{e,\parallel}$ and $g_{h,\parallel}$, which are the electron and effective heavy-hole magnetic $g$-factors, and $\Delta_i$ ($i=0,1,2)$, which represent the short-range electron-hole exchange interaction, are functions of the QW width \cite{kesteren,snelling} (see section \ref{sec:11} and Appendix I). Note that the expression \ref{equ6} is valid down to $C_{2v}$ symmetry.

The energies of the four heavy-hole exciton states for applied field parallel to $Oz$ are:
\begin{subequations}
\begin{equation}
	\delta E_{\pm1}=\frac{\Delta_0}{2}\mp\frac{1}{2}\sqrt{ \mu_B^2 B_z^2(g_{h,\parallel}+g_{e,\parallel})^2 +\Delta^2_1 }
	\label{equ7a}
	\end{equation}
\begin{equation}
  \delta E_{\pm2}=-\frac{\Delta_0}{2}\mp\frac{1}{2}\sqrt{ \mu_B^2 B_z^2(g_{h,\parallel}-g_{e,\parallel})^2 +\Delta^2_2 }
	\label{equ7b}	
\end{equation}
\label{equ7}
\end{subequations} 
The levels are plotted in figure \ref{Fig15} for the ideal $D_{2d}$ symmetry and for $\Delta_1<<\Delta_0$ \cite{blackwood}. The $z$ component of exchange ($\Delta_0$) causes a zero-field splitting between the optically allowed and nonallowed states and the $\Delta_1$ and $\Delta_2$ components cause small additional zero-field splittings. $D_{2d}$  has a fourfold rotation-reflection axis along the growth direction (\textit{Oz}) which dictates $\Delta_1=0$, so that $E_{+1}$ and $E_{-1}$ are degenerate in zero field. If this symmetry is broken a zero-field splitting $\Delta_1$ appears (see section \ref{sec:5} on type II quantum wells).

The two fields at which the exciton levels cross are given by :
\begin{eqnarray}
	B_z^{(h)}\approx\frac{\Delta_0}{g_{h,\parallel}\mu_B} \hspace{0.5cm} and  \hspace{0.5cm} 	B_z^{(e)}\approx\frac{\Delta_0}{g_{e,\parallel}\mu_B}
\end{eqnarray}
\begin{figure}[t]
	\centering
		\includegraphics*[width=.7\textwidth]{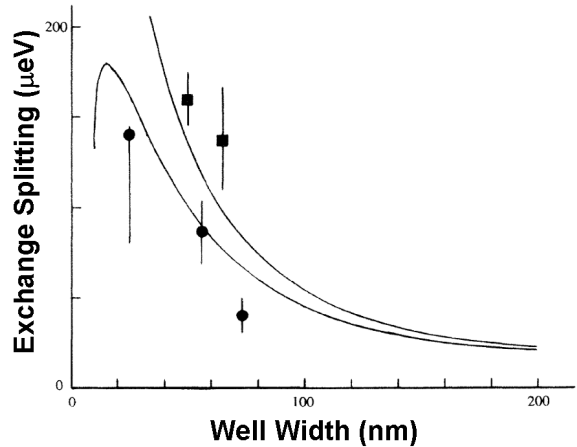}
		\caption{Exciton exchange energy  $\Delta_0$  as a function of the well widths in GaAs/AlGaAs ($\bullet$) and GaAs/AlAs ($\scriptstyle{\blacksquare}$) QW. The full lines are the calculated values  \cite{blackwood}.}
	\label{Fig16}
\end{figure}
From the measurements of the electron and hole $g$ factors \cite{snelling,snelling2}, it turns out that $\left|g_{h,\parallel}\right|>\left|g_{e,\parallel}\right|$. The peaks observed in figure \ref{Fig14} is thus associated to $B_z^{(h)}$ ($B_z^{(e)}$ is beyond the range of measurement). The measurement of $B_z^{(h)}$ in figure 	\ref{Fig14} thus leads to the value of the exciton exchange energy $\Delta_0$ plotted in figure \ref{Fig16}. The exchange energy increases rapidly as the QW width decreases and as the barrier height increases. The values are in satisfactory agreement with calculations of the enhancement of the exchange relative to the bulk value ($\approx10 \pm5\mu eV$) due to enhanced electron-hole overlap \cite{blackwood}, as expected from section \ref{sec:11}.

\subsubsection{b) Exciton g-factor}
\label{sec:412}

\begin{figure}[t]
	\centering
		\includegraphics*[width=.7\textwidth]{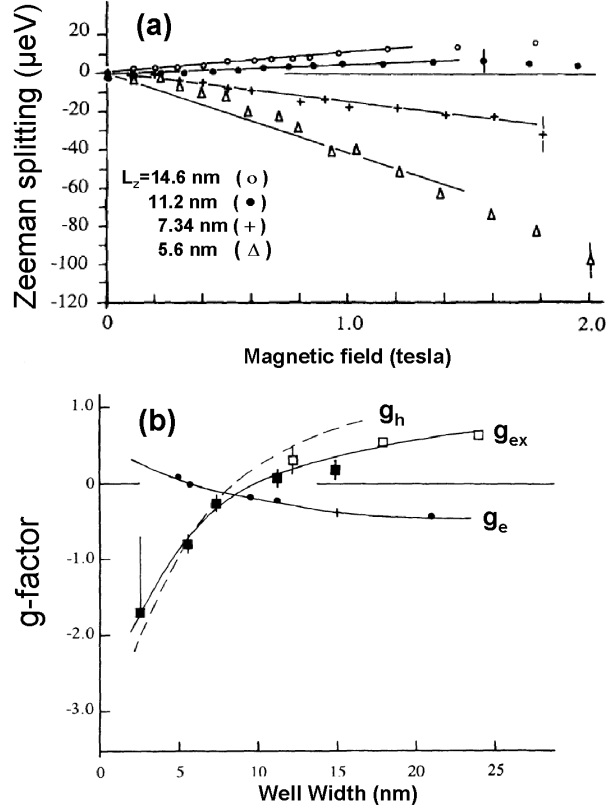}
		\caption{(a)	Low-field Zeeman splitting of the XH exciton luminescence lines in GaAs/AlGaAs QW at $T=1.8 K$
(b)	Electron ($g_e\equiv g_{e,\parallel}$), heavy-hole ($g_{h}\equiv g_{h,\parallel}$) and exciton ($g_{exc}\equiv g^1_{exc,\parallel}$) $g$-factors in GaAs/AlGaAs QW \cite{snelling}.}
	\label{Fig17}
\end{figure}

The magnetic g-factor for the heavy-hole exciton in GaAs/AlGaAs QW has been determined as a function of well width from the Zeeman splitting of the cw luminescence spectra for moderate longitudinal magnetic fields (to avoid level crossings presented above) \cite{snelling}. Figure \ref{Fig17}(a) shows the measured Zeeman splittings up to $B_z=2 T$ for different well widths. The variations as a function of the magnetic field are linear within the experimental uncertainties and the slopes give the values of $g^{J=1}_{exc,\parallel}=g_{e,\parallel}+g_{h,\parallel}$ which are plotted in figure 	\ref{Fig17}.b, showing the change of sign for $L_W$ between $7$ and $11 nm$. 
%

\subsection{ Exciton spin Quantum Beats spectroscopy }\label{sec:42}

Thanks to the development of ultra-fast lasers and sensitive detectors, it has been possible to measure in the time domain the interaction of the exciton states with the external magnetic field \cite{barad,dda,amand2,worsley,dyakonov}. This leads to measurements with a great accuracy of the exciton $g$-factor and exciton exchange energy.

The principle is the following. When two energetically closely spaced transitions are excited with a short optical pulse (with a spectral width larger than the splitting between the transitions), the two-induced polarizations in the medium oscillate with their slightly different frequencies. Their interference manifests itself in a modulation of the net polarization, the so-called Quantum Beats (QB) \cite{haroche}. This allows energy splittings to be determined with higher resolution than in the spectral domain, provided that the beats period is shorter than their damping.

\subsubsection{a) Exciton spin Quantum Beats in longitudinal magnetic fields }
\label{sec421}

The exciton spin dynamics in longitudinal magnetic field (applied along the QW growth axis, Faraday configuration) has been measured with different experimental techniques, including time-resolved pump-probe transmission \cite{barad}, time-resolved Faraday Rotation \cite{baumberg,dda,worsley} and time-resolved photoluminescence \cite{amand2}.

In a longitudinal magnetic field, the optically active exciton states are the $\left|+1\right\rangle$ and $\left|-1\right\rangle$ states split by the Zeeman energy $\hbar\Omega_\parallel=g_{exc}\mu_BB_z$,  where $g_{exc}=g_{e,\parallel}+g_{h,\parallel}$. A linearly-polarized optical excitation pulse, resonant with the exciton energy, will thus create a coherent superposition of $\left|+1\right\rangle$ and $\left|-1\right\rangle$ states, making the observation of quantum beats as a function of time possible.
\begin{figure}[t]
	\centering
		\includegraphics*[width=.7\textwidth]{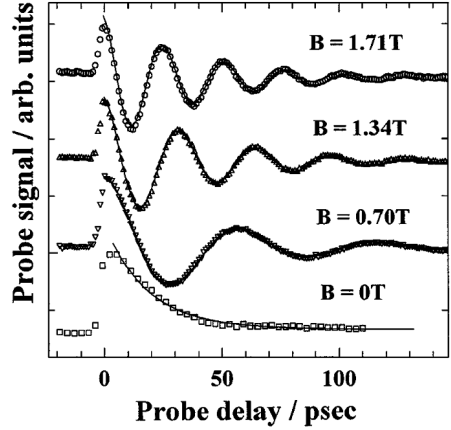}
		\caption{Quantum beats observed in the transient birefringence from the XH exciton in a $2.75 nm$ GaAs/AlGaAs QW at $T=1.8 K$ for various applied longitudinal magnetic fields \cite{worsley}.}
	\label{Fig18}
\end{figure}
Figure 	\ref{Fig18} shows the transient birefringence from the heavy-hole exciton in a $2.75 nm$ GaAs/AlGaAs MQW structure for various longitudinal magnetic fields \cite{worsley}. The pump pulse (linearly-polarized) is resonant with the exciton absorption and the probe pulse has a linear polarization tilted by an angle of $45^\circ$ with respect to the pump polarization. In this time-resolved Kerr rotation experiment, the transient pump-induced birefringence plotted in figure \ref{Fig18} corresponds to the degree of induced elliptization of the probe pulse reflected from the sample. At zero field, there is an exponential decay, which corresponds to the coherent decay of the exciton linear polarization ; the corresponding decay time $T_{s2}^{*}$ is given by $1/T_{s2}^{*}=1/T_{s2}+1/\tau_{rad}$ where $T_{s2}$ is the so-called exciton transverse spin relaxation time presented in \ref{sec:34}a and $\tau_{rad}$ is the radiative lifetime \cite{mas}. As the magnetic field increases, the QB observed in figure \ref{Fig18} correspond to the coherent oscillation between the Zeeman-split exciton levels ($M=\pm1$) at the pulsation $\Omega_\parallel= g_{exc}\mu_B B_z/\hbar$. The fit of the data in figure \ref{Fig18} gives the Zeeman splitting from which the exciton $g$ factor $\left|g_{exc}\right|=1.52\pm0,01$ is obtained for a $L_W=2.75 nm$ MQW. This measurement is much more accurate than the ones performed previously in the spectral domain presented in figure \ref{Fig17}b \cite{snelling}.

\subsubsection{b) Exciton spin Quantum Beats in transverse magnetic fields }\label{422}

The spin Hamiltonian of the heavy-hole exciton in a transverse magnetic field (applied in the QW plane, $\bm B//Ox$) deduced from \ref{sec:11} and \ref{sec:13} can be approximated by :
\begin{eqnarray}
	\bm H=\hbar \omega \bm S-\frac{2\Delta_0}{3} J_zS_z
\label{equ13}	
\end{eqnarray} 
where $\hbar\omega=g_{e,x}\mu_BB_x$  and  $\Delta_0$ is the zero-field exciton exchange splitting between the optically active states $\left|\pm1\right\rangle$ and the two dark states $\left|\pm2\right\rangle$ (the much smaller splitting between the $\left|+2\right\rangle$ and $\left|-2\right\rangle$ states is neglected, as well as $\Delta_1$) \cite{blackwood,ivchenko}. We assume here that the transverse $g$-factor of the $j_{h,z}=\pm3/2$ heavy hole is zero (Spin QB experiments performed in n-doped GaAs QW show that $g_{h,x}\approx 0.04$ \cite{mariePRB}, so that $g_{h,x}<<g_{e,x}=g_{e,y}$).

\begin{figure}[t]
	\centering
		\includegraphics*[width=.7\textwidth]{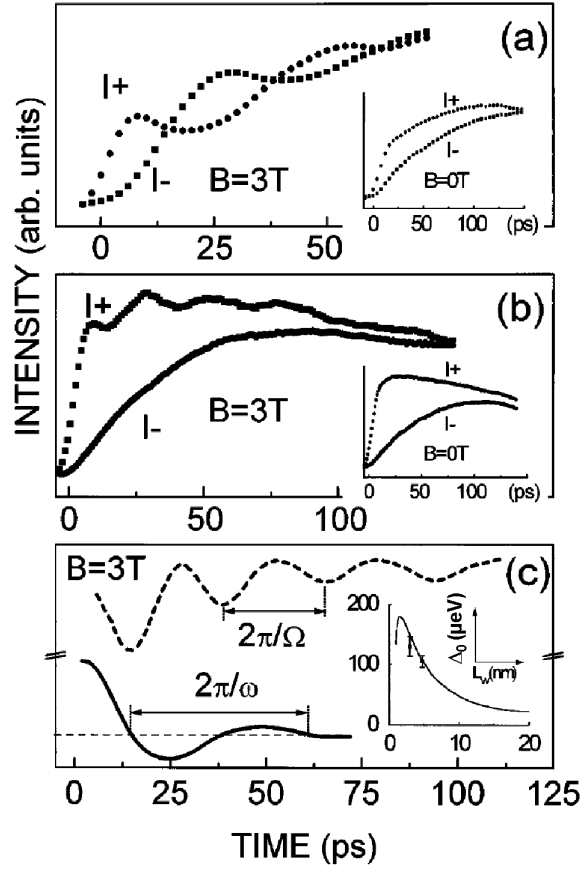}
		\caption{Luminescence intensity dynamics after $\sigma^+$ polarized excitation in a $L_w=3nm$ GaAs MQW, at T=1.7 K. (a) The excitation energy is non-resonant ($E_1-HH_1 < h \nu <  XL$,XL is the light-hole exciton energy)  and $B =3 T$ (inset, $B = 0 T$), (b) The excitation energy is resonant with XH and $B =3 T$ (inset, $B = 0 T$). (c) The oscillations of the luminescence intensity component $I^+$ in resonant excitation (dashed line) and of the luminescence polarization PL in non-resonant excitation $E_1-HH_1 < h \nu <  XL$ (full line), under the same magnetic field $B =3 T$. For the sake of clarity, the monotonous component has been subtracted from $I^+$. Inset:  well-width dependence of the exciton exchange energy $\delta_0$, from this experiment (dots with error bars) and theory \cite{blackwood,amand2} (full line).}
	\label{Fig19}
\end{figure}

The exciton quasi-stationary states $\left|\Psi_+\right\rangle$ in the transverse magnetic field are two linear combinations $E_\pm$ of optically and inactive states split by the energy $\hbar\Omega_{exc}$  and write \cite{amand2,mashkov,puls} :
\begin{eqnarray}
	\left|\Psi_+\right\rangle&\approx&\hbar\omega\left|1\right\rangle+(\hbar\Omega_{exc}-\Delta_0)\left|2\right\rangle\\
		\left|\Psi_-\right\rangle&\approx&-(\hbar\Omega_{exc}-\Delta_0)\left|1\right\rangle+\hbar\omega\left|2\right\rangle
\end{eqnarray}
where $\hbar\Omega_{exc}=\sqrt{\Delta_0^2+(\hbar \omega)^2}$, and $\omega=\delta_{e,B}/\hbar$.
A ($\sigma^+)$-polarized pulsed excitation resonant with the exciton energy will thus create a coherent superposition of  $\left|\Psi_+\right\rangle$ and  $\left|\Psi_-\right\rangle$ states. Ignoring any spin relaxation processes (as well as recombination), the right ($I^+$) and left ($I^-$) circularly-polarized luminescence components are proportional to $\left|\left\langle \pm1|\Psi(t)\right\rangle\right|^2$ :
\begin{eqnarray}
	I^+(t)&=&1-\left(\frac{\omega}{\Omega_{exc}}\right)^2\left(\frac{1-cos(\Omega_{exc}t)}{2}\right)\\
	I^-(t)&=&0
\end{eqnarray}
As a consequence, we expect to observe, in time-resolved photoluminescence, oscillations of the polarized emission $I^+(t)$ which should occur with a pulsation  $\Omega_{exc}$, \textit{i.e.} the pulsation should not depend linearly on  the applied transverse field. The co-polarized luminescence intensity $I^+$, modulated at the pulsation  $\Omega_{exc}$ has an amplitude reduced by a factor $(\omega/\Omega_{exc})^2$, while the counter-polarized component is unmodulated in this simplified approach.  These exciton-like spin QB were indeed observed in narrow MQW samples. Figure \ref{Fig19}.b  presents the luminescence intensity dynamics co-polarized ($I^+$) and counter-polarized ($I^-$) with the resonant ($\sigma^+$)-polarized pisosecond laser in a $L_W=3 nm$ MQW GaAs/AlGaAs sample. Quantum Beats are observed only at strong magnetic field values ; they appear as a weak amplitude modulation on the $I^+$  component but are not observable on $I^-$ \cite{oestreich,gerlovin}.
If the excitation energy is higher than the QW band gap ($E_1-HH_1$), all the MQW samples exhibit QB on $I^+$ and $I^-$ with an oscillation frequency proportional to the magnetic field, see fig.\ref{Fig19}a \cite{heberle,hannak}. Oscillations on $I^+$ and $I^-$ are phase shifted by $\pi$. These oscillations are attributed to the Larmor precession of the free electron with pulsation $\omega$. This yields the accurate measurement of $g_{e,x}=0.50\pm0.01$ in figure \ref{Fig19}.a. When the laser excitation is resonant, we see clearly in figure \ref{Fig19}.c that the beat period is very different than in the non resonant case. It is attributed to the exciton QB and can be used to measure the exciton exchange energy $\Delta_0=\hbar\sqrt{\Omega^2_{exc}-\omega^2}$; $\Delta_0=130\pm15$ $\mu$eV and $\Delta_0=105\pm10$ $\mu$eV are measured in a $L_W=3 nm$ and $L_W=4.8 nm$ GaAs/AlGaAs MQW structure respectively \cite{amand2}.

The following question arises now: why in most of the experiments performed in transverse magnetic fields do the authors observe electron QB (with a pulsation  $\omega=g_{e,\bot}\mu_BB/\hbar$) and not the exciton QB (with a pulsation  $\Omega_{exc}=\hbar^{-1}\sqrt{\Delta_0^2+(\hbar\omega)^2}$) though the recorded signal corresponds to exciton transitions \cite{heberle,dda}. This enigma has been explained by D'Yakonov \textit{et al.} \cite{dyakonov}. It turns out that the observation of QBs on the excitonic luminescence at the electronic or excitonic pulsation ($\omega$ or $\Omega_{exc}$ respectively) is related to the stability of the
hole-spin orientation within the exciton. The argument is the following. Within the exciton, the correlation between electron and hole spins is held by the electron-hole exchange interaction. However, if this correlation is not strong enough to reduce the single-particle hole spin flip at a rate lower than  $\Delta_0/\hbar$, the exchange interaction splitting  $\Delta_0$  no longer plays a
role in the QB. Then the QB appears at the pulsation $\omega$. Finally an electron bound into an exciton precesses like a free electron in the transverse magnetic field provided that $\tau_h<<\hbar/\Delta_0$ where $\tau_h$ is the single-particle hole spin-flip time. This condition can be fulfilled in large and narrow QW's but for different reasons. In large QWs (\textit{a fortiori }in  bulk material)  such a hole spin flip occurs as a consequence of the mixing of states in the valence band due to spin-orbit interaction and small exchange interaction; the observation of the electron precession in a QW of $25 nm$ well width \textit{under the resonant or non-resonant} excitation conditions reported initially by Heberle\textit{et al.}  is understood on this ground \cite{heberle}. In narrow QWs the hole spin flip, which results in the observation of QB's at the pulsation $\omega$  in non-resonant excitation, is related to the formation-dissociation process of excitons and the related long cooling of the excited system \cite{robart,deveaud}. QBs of the excitonic kind have been observed only in narrow quantum wells ($L_W<10 nm$) \textit{under resonant excitation}. This indicates again that the hole-spin orientation is rather stable in cold two-dimensional excitons ($\tau_h>\hbar/\Delta_0$), see section \ref{sec:32}. We give in Appendix II a simple explanation of exciton splitting in transverse magnetic field with unstable hole spin in terms of exchange coupling strength between the electron and the hole spins. A model based on a density matrix approach has been developped in ref. \cite{dyakonov}, which can reproduce the characteristic experimental features described above.

\section{Exciton spin dynamics in type II QWs }
\label{sec:5}

In the previous sections, we discussed the exciton spin properties in the so-called type I quantum well structures, \textit{i.e.} where the conduction electron and the valence holes are confined in the same material and the same region in space. In $GaAs/AlGaAs$ QWs depending on the well width and $Al$ percentage it turns out that two types of lowest energy transitions are possible. For low $Al$ content the conduction band - confined states in the well has the lowest energy (type I quantum well). For large $Al$ content and small well width, the lowest conduction band - confined state in the well has a higher energy than the lowest $X$-confined state in the barrier \cite{dawson}. As the holes are still confined in the $GaAs$ well, the recombination takes therefore place between the hole in the well and the electrons in the barrier (type II quantum well). This accounts for the long exciton PL lifetime, of the order of a few microseconds \cite{vanderpoel2}.
\begin{figure}[t]
	\centering
		\includegraphics*[width=.7\textwidth]{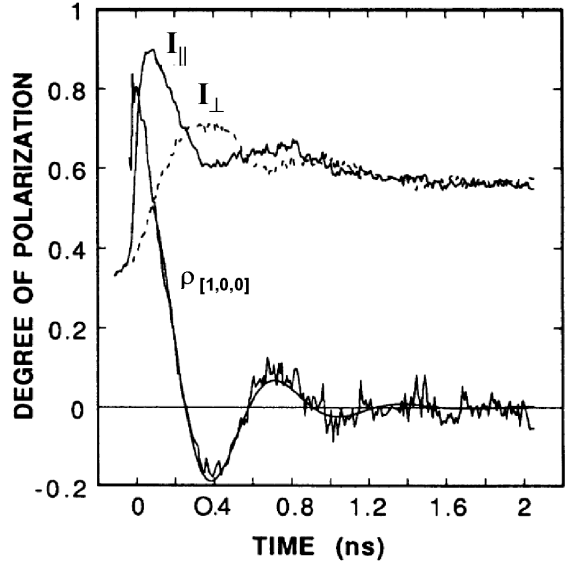}
		\caption{Type II GaAs/AlAs 2.2/1.15 nm superlattice; $T=4.2 K$.
Degree of PL linear polarization  $\rho_{[1,0,0]}$ as a function of time. The excitation picosecond laser pulse is linearly-polarized along the $[1,0,0]$ axis. The intensities $I_{//}$ and $I_\perp$ are detected with polarization parallel and perpendicular to the excitation respectively \cite{gourdon}.}
	\label{Fig20}
\end{figure}

The spin dynamics in these type II quantum well systems has been extensively studied by optical orientation experiments in stationary or time-resolved regime \cite{vanderpoel,kesteren,gourdon,mashkov,ivchenko}. Because of the very small overlap between the electron and hole wavefunction, the spin relaxation mechanism induced by the exchange interaction between the electron and the hole does not play a significant role in contrast to type I QWs (see sections \ref{sec:12}, \ref{sec:34}). However the strong localization of the carrier wavefunction at the QW interface will (\textit{i}) modify drastically the exciton fine structure and (\textit{ii}) yield very long electron spin relaxation times ($\approx$ a few tens of ns) compared to type I QWs \cite{silva}. It has been shown that the symmetry of the system is reduced form $D_{2d}$ to $C_{2v}$ and the two optically active exciton eigenstates are linearly-polarized, split by an energy of a few $\mu eV$ and aligned along the $X'\equiv[1,1,0]$ and $Y'\equiv[1,-1,0]$ crystal directions \cite{kesteren,vanderpoel,gourdon}. The symmetry reduction in these type II GaAs superlattices was first explained by the presence of a random local deformation (due to the presence of bonds of different nature (Al-As or Ga-As) on each side of the interface taken as an As plane) which mixes the heavy and light hole states \cite{gourdon}.  It was then shown that the splitting, called anisotropic exchange splitting, arises from intrinsic effect: the mixing of heavy- and light-hole states at the interface due to the low ($C_{2v}$) symmetry of the interface.
Since the splitting of the X' and Y' excitonic sublevels (which is much smaller than $k_BT$), is much smaller than the laser spectral width, the two sublevels can be coherently excited at time $t=0$ by a short pulse polarized along one of the $[1,0,0]$ axes (\textit{i.e.} 45$^\circ$ angle with respect to the exciton eigenstates orientations) \cite{vanderpoel}. As a consequence, the time-resolved luminescence signal, detected with polarization either parallel or perpendicular to the excitation, decays and oscillates with a period $T$ inversely proportional to the splitting $\Delta_1$ between the two sublevels ($T=h/\Delta_1$, see section \ref{sec:11}). The time dependence of the photoluminescence linear polarization is shown in figure \ref{Fig20} for a $2.2/1.5 nm$ type II GaAs/AlAs superlattice \cite{gourdon}. The period $T$ is about $640 ps$, which corresponds to an energy splitting of $\approx 6.3 \mu eV$. 
The application of a transverse magnetic field leads to a complex oscillation pattern with several frequencies, since, besides the coupling between  $\left|+1\right\rangle$ and  $\left|+2\right\rangle$ and  $\left|-1\right\rangle$ and  $\left|-2\right\rangle$ exciton states induced by the external magnetic field, the anisotropic exchange also couples the  $\left|+1\right\rangle$ and  $\left|-1\right\rangle$ states \cite{mashkov}. 

\section{Spin dynamics in dense excitonic systems}
\label{sec:6}

When the areal exciton density becomes non negligible with respect to the critical density defined by  $n_c=$\begin{small}$\left[32\pi\left(a^{2D}_B\right)^2\right]^{-1}$\end{small} (where $a^{2D}_B\equiv a^{3D}_B/4$ is the two dimensional Bohr radius), the exciton mutual interactions start to modify significantly the single exciton picture we used up to now. The above mentioned critical density $n_c$ corresponds to the one where the exciton binding energy is zero, due to phase space filling and screening of the Coulomb interaction. If the areal exciton density $n_{ex}$ approaches $n_c$ ($n_{ex} \lesssim n_c$), the exciton energies $E_{\bm K}$ must be corrected by a complex self energy term, which real part corresponds to the energy shift and the imaginary part to the broadening of the single exciton states due to the mutual Coulomb interactions \cite{Mahan}. We will show in the two following sections the experimental manifestations of these two complementary aspects in the case of two-dimensional structures, namely the spin dependent exciton energy shift and the spin dependent exciton-exciton collisions at high exciton densities, as revealed in elliptically polarized exciton populations. 
In the case of excitation by elliptical light, excitons are created in the elliptical states:
\begin{eqnarray}
	\left|E_{\theta}\right\rangle = sin\left(\theta + \pi/4 \right)\left|+1\right\rangle +cos\left(\theta + \pi/4 \right)\left|-1\right\rangle
\label{equation61}
\end{eqnarray}            
so that linear excitons are given by $\left|X\right\rangle=\left|E_0\right\rangle$  and  $i\left|Y\right\rangle=\left|E_{\pi/2}\right\rangle$. Excitons $\left|+1\right\rangle=\left|E_{\pi/4}\right\rangle$ 
and $\left|-1\right\rangle=\left|E_{-\pi/4}\right\rangle$ , excited by $\sigma^+$ or $\sigma^-$  light respectively, are called circular excitons. The circular polarisation of the state $\left|E_{\theta}\right\rangle$ is simply $P_c(\theta) = sin(2\theta)$, while the linear one is  $P_c(\theta) = cos(2\theta)$.
We are interested here to describe experiences performed at low temperature and under resonant (or quasi-resonant) excitation with the heavy-hole excitons. The latter are thus created in the 1s state with very small, or even zero wave vector, \textit{i.e.} with very small kinetic energy. As a consequence, the scattering probability to 2s, or 2p states, which are close to the QW gap in 2D systems (see Appendix I), is low. We restrain thus to the Heavy-Hole exciton subspace, and choose the basis $\mathcal{B}_{XH}=\left\{ \left|\right.\pm2\left.\right\rangle, \left|\right.\pm1\left.\right\rangle \right\}$. The general form of exciton-exciton interaction is recalled in Appendix III. The final result is that the interaction Hamiltonian can be approximated for an exciton pair (\textit{i},\textit{j}), in a cold exciton population with low density, as:
\begin{eqnarray}
H^{i,j}_{exch}(\bm K,\bm K',\bm Q)\approx \frac{6e^2a^{2D}_B}{\epsilon_0 A}\left( \bm {\sigma}^{(i)}_e \bm . \bm{\sigma}^{(j)}_e	+ 
\bm {\sigma}^{(i)}_h \bm . \bm{\sigma}^{(j)}_h + 2 \right) 
\label{equation62}
\end{eqnarray}    
where : $\bm {\sigma}^{(i)}_{e(h)}\equiv \left(\right.\sigma^{(i)}_{e(h),x}, \sigma^{(i)}_{e(h),y},\sigma^{(i)}_{e(h),z}\left.\right)$ represent Pauli matrices vector operators for electron spins and heavy-hole effective spins, and $A$ is the QW quantization area. This expression results from the fact that the electron-electron or the hole-hole exchange dominate over direct Coulomb interaction as well as exciton exchange as a whole, provided that the initial wave vector $\bm K$ and $\bm K'$ of the excitons which interact are small with respect to $(a^{2D}_B)^{-1}$ (see Appendix III). In such conditions, it neither depends on $\bm K$, $\bm K'$, nor on the  wavevector $\bm Q$  transferred during the collision.

\subsection{Exciton spin-dependent renormalization}
\label{sec:61}

First evidence of exciton state renormalisation at high density were obtained by D. Hulin \textit{et al.}, who observed, in femtosecond pump-probe experiments performed on GaAs/AlGaAs multiquantum well structures under linearly polarized pump, a transient blue-shift of the exciton absorption line \cite{Peyghambarian, Hulin}. The latter was shown to be tied to the reduced dimensionality of excitons, being well apparent in GaAs wells of thickness of the order of 5 nm, but disappearing rapidly for larger well sizes. The authors interpreted this effect in terms of a strong reduction of long-range many-body interactions in a 2D system, in agreement with the theory of Schmitt-Rink \textit{et al.} \cite{Schmitt-Rink}. It is well documented that in 3D systems, the exciton absolute energy remains unchanged, even at high densities \cite{Haug}. This energy constancy is attributed to the almost exact compensation between two many-body effects acting in opposite directions: an inter-particle attraction which, for bound electron-hole pairs at $T\approx 0$, is similar to a van der Waals interaction, and a repulsive contribution having its origin in the Pauli exclusion principle acting on the Fermi particles (electron and holes) forming the excitons. The argument of Schmitt-Rink \textit{et al.} is that the long-range attractive component is strongly reduced in a 2D system, so that the short-range repulsive part becomes now unbalanced.  

\begin{figure}[t]
	\centering
		\includegraphics*[width=.8\textwidth]{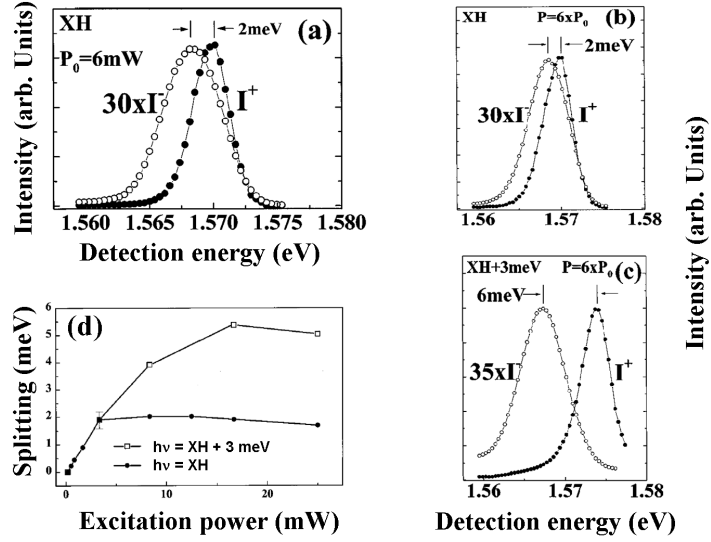}
		\caption{Spectra of the exciton luminescence components co-polarized ($I^+$) and counter-polarized ($I^-$) with the circularly polarized excitation ($P_E\approx 1$) at time delay $t=4 ps$. (a) the excitation energy is set to $h\nu_E= XH$. The average excitation power $P$ is $P0 = 6 mW$. (b) $h\nu = XH$ and $P = 6P_0$ (c) $h\nu = XH +3 meV$ and $P = 6P_0$. (d) Splitting energy between the two luminescence components $I^+$ and $I^-$ as a function of $P$ : ($\bullet$) $h\nu = XH$ ; ($\scriptstyle{\Box}$) $h\nu = XH +3 meV$ \cite{LeJeune2}.}
	\label{Figure2}
\end{figure}

Using now circularly polarized excitation in time resolved polarised luminescence, it has been shown that this blue shift was spin dependent, and that in a dense and circularly polarized exciton gas, a splitting occurs between the line co-polarised with the excitation, and the counter-polarised one, the former experiencing a blue shift, while the latter is red shifted \cite{damen2,dareys2,Amand3}. We show here on figure \ref{Figure2} the result of an experiment performed under resonant excitation on a high quality $GaAs/AlGaAs$ multi quantum well structure (i.e. with Stokes shift less than 0.1 meV and cw PL line width $\Gamma\approx0.9 meV$ at $T\approx1.7 K$). Two-colour time resolved up-conversion photoluminescence spectroscopy was used to perform such experiments, the excitation laser pulse duration being $\delta t\approx1.5 ps$, so that only the 1s heavy-hole exciton state is excited \cite{LeJeune2}. Just after a $\sigma^+$ circularly polarized excitation resonant with the XH  exciton ($P_E\approx1$, $h\nu_E = XH$), the strongly polarized emission ($P_L\approx0.9$) displays a splitting between the co-polarised emission line $I^+$ and the counter polarized one $I^-$.  

In addition, the $I^+$ emission is strongly blue shifted, while the $I^-$ is slightly red shifted. When increasing the
 excitation power, the energetic positions first vary linearly. A saturation then occurs at $P_{sat}\approx3mW$, when the $I^+$ line shifts to energies higher than the XH energy by the laser line width ($\delta_E\approx2meV$). The absorption then
  drops, due to energy mismatch between the laser and the renormalized XH exciton energy. This situation corresponds to the exciton density $n_{sat}$ estimated at $n_{sat}\sim 2\times 10^{10}cm^{-2} $. The splitting amounts then to about
   $\delta E_{+1}\approx  1.9 meV$. For $P > Psat$, a self regulation effect appears for the exciton density. When the excitation energy is increased at $XH+3meV$, the saturation effect occurs at higher excitation power. The blue shift can
    become a significant fraction of the 1s exciton binding energy, here estimated at $E_B\approx 8 meV$. The saturation exciton density is below the critical density $n_c$, here estimated to about $3\times10^{11} cm^{-2}$. A good
     phenomenological description of the line positions in the linear density regime ($n_{ex}<<n_c$) can be obtained for a cold exciton gas, according to \cite{Amand3}:
\begin{subequations}
\begin{equation}
	\delta E_{\pm1}=K_1 n_{\pm1}+\frac{1}{2}K_1(n_{+2}+n_{-2})-K_2 n_{ex}
	\label{equation63a}
	\end{equation}
\begin{equation}
  \Delta E\equiv E_{+1}-E_{-1}=K_1 \left( n_{+1}-n_{-1}\right)
	\label{equation63b}	
\end{equation}
\label{equation63}
\end{subequations} 
where the $n_M$ are the exciton population densities corresponding to the $\left|M\right\rangle$ states ($n_M=N_M/A$), $n_{ex}$ is the
 total exciton density, and $K_1$ and $K_2$ are positive constants. The first one, $K_1$, represents the strength of the
  repulsive part of the interaction between $J=1$ excitons of the \textit{same} angular momentum projection $M$, which takes
   its origin in the Pauli repulsion principle. Since the $\left|\left. \right.+1\right\rangle$ exciton shares its
    electron spin states with the $\left|-2\right\rangle$ exciton, but not its hole state, the contribution to the
     exchange energy shift of $\left|+1\right\rangle$ excitons by the $\left|-2\right\rangle$ excitons is taken as 
     $(K_1 /2)n_{-2}$. A similar reason holds for the contribution of $n_{+2}$ excitons to $\delta E_{+1}$. The constant $K_2$, also positive, represents the weak attractive part of the interaction between excitons; for the sake of
      simplicity, it is assumed spin independent \cite{Schmitt-Rink, Fernandez-Rossier}. The experimental values of
       $K_1$ and $K_2$ can be determined from the initial splitting $\Delta E$ and energy shift  $E_{-1}$ just after the
        resonant excitation by $\sigma^+$ light, so that $n_M(0)=n_{ex}(0)\delta_{M,+1}$.
   It is found that $10^{-10} \lesssim K_1 \lesssim 1.6 \times 10^{-10} meV cm^2$, depending on  the quantum well, and 
         $K_2/K_1\sim 0.15$ typically. From the theoretical calculation of Schmitt-Rink \textit{et al.} for a non-polarized
          exciton gas \cite{Schmitt-Rink}, we can infer that :
           $K_1\approx2\times3.86\pi\left(a^{2D}_B\right)^2E^{2D}_B \approx 4\times 6.06\frac{e^2}{2\epsilon_0 a^{2D}_B}$\footnote{the theoretical value of $K_1$ can also be found from a description of the exciton gas in term of interacting quasi-bosons. The exchange shift constant is then approximated to $K_1\approx 4 \times 6.0 \frac{e^2}{2\epsilon_0} a^{2D}_B$ \cite{Ciuti-Schwendimann}}.  
Considering now $a^{2D}_B$ as a variational parameter $a_{eff}$ , which is a good approximation for narrow QWs with marked 2D character \cite{Bastard2}, (note: the relation $E_B a_{eff} =e^2/(2\epsilon_0)$ interpolates correctly between the 2D and the 3D case), the theoretical estimation of $K_1$ is in reasonable agreement with the experimental values. 

\subsection{Exciton spin dynamics under elliptical optical pumping: exchange assisted transfer between dark and bright states}
\label{sec:62}

\begin{figure}[t]
	\centering
		\includegraphics*[width=.8\textwidth]{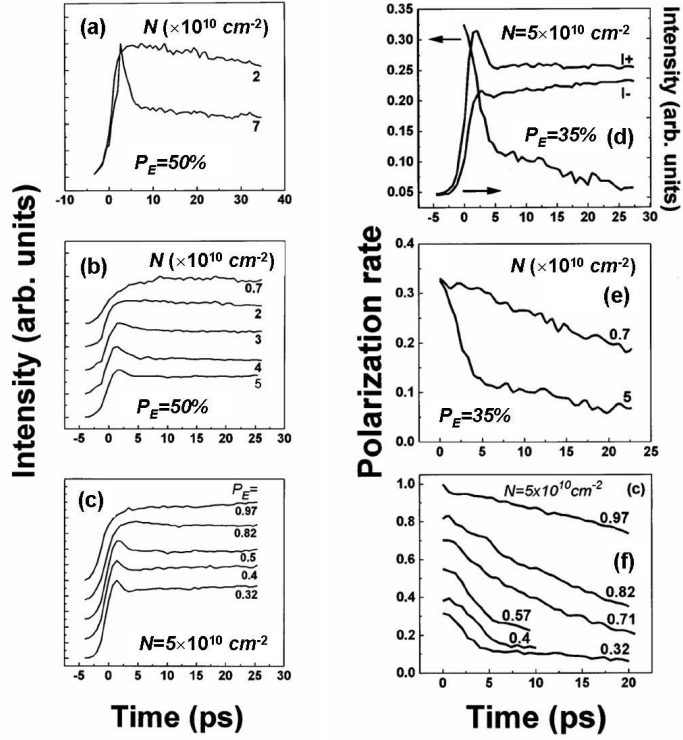}
		\caption{Normalized total luminescence intensity under elliptical excitation with circular polarization $P_E=0.5$, at (a): $n(0) \sim 2\times 10^{10}$ and $7\times 10^{10} cm^{-2}$;  (b): with $n(0)$ ranging from $7\times 10^9 cm^{-2}$ to $5\times 10^{10} cm^{-2}$. (c) Normalized total luminescence intensity at $n(0)\sim 5\times 10^{10} cm^{-2}$ with $P_E$ ranging from $0.32$ to $0.97$. (d): Luminescence intensities $I^+(t)$ and $I^-(t)$, and circular polarization $P_L(t)$ at $n(0)\sim 5\times 10^{10} cm^{-2}$ and $P_E= 0.35$; (e): time evolution of $P_L(t)$ at $n(0) = 7\times 10^9 cm^{-2}$ and $5\times 10^{10} cm^{-2}$; (f) $P_L(t)$ at $n(0)\sim 5\times 10^{10} cm^{-2}$ for $P_E$ ranging from $0.35$ to $0.97$ \cite{Amand-Robart}.}
	\label{Figure3}
\end{figure}

Besides its contribution to renormalisation, the exchange interaction between excitons at high density is at the origin of specific exciton spin relaxation processes. This is first apparent in the emission line broadening, where a strong asymmetry is observed (cf. figure \ref{Figure2}). The luminescence component co-polarized with the laser excitation ($I^+$) is narrower than the counter polarized one ($I^-$), although $n_{+1} >> n_{-1}$. The $n_{-1}$ population arises from the small fraction of excitons which have lost their initial polarization at $t=4 ps$, the circular polarization of the emission being $P_L=0.9$. The new spin relaxation processes appearing at high density are illustrated in the figure  \ref{Figure3}, which shows time resolved photoluminescence results performed on a 60 periods $Al_{0.3}Ga_{0.7}As$ multi-quantum well structure grown on a [1,0,0] substrate, excited resonantly with the XH exciton, and presenting a Stokes shift between the absorption and the emission of 6 meV.   

The salient features of dense and elliptically polarized exciton gas are :
(\textit{i}) fast decay of the total emission intensity, followed by a much slower one. The fast component disappears in the low-excitation regime and in the limiting case of pure circular polarization at any density. (\textit{ii}) fast decay of the circular polarization of the emission, which is correlated to the intensity decay, and which disappears also at low density or when a pure circular exciton population is generated. The initial decay time of the PL circular polarization decreases, when the ellipticity is increased, down to values much shorter than the low density exciton longitudinal spin relaxation time $T_{s1}$. For a pure circular exciton population, the circular polarization dynamics does not depend on the exciton density. All these characteristics have also been observed in samples without Stokes shift \cite{LeJeune2}. 

The detailed interpretation has been given in ref. \cite{Amand-Robart}. In a dense gas of elliptical excitons, the exchange interaction between excitons becomes stronger than the internal electron-hole exchange within single excitons, and destroys the intra-exciton spin coherence. The equations showing the action of the exchange hamiltonian $H_{exch}$ in elliptical and circular cases, as deduced from equation (\ref{equation62}), are:
\begin{subequations}
\begin{equation}
	H_{exch}\left|+1\right\rangle\left|+1\right\rangle = \frac{K_1}{A}\left|+1\right\rangle\left|+1\right\rangle
	\label{equation64a}
	\end{equation}
\begin{eqnarray}
  H\left|E_{\theta}\right\rangle\left|E_{\theta}\right\rangle=  
    \frac{K_1}{A}\left[\frac{1+sin2\theta}{2}\left|+1\right\rangle\left|+1\right\rangle
  + \frac{1-sin2\theta}{2}\left|-1\right\rangle\left|-1\right\rangle \right.
  \nonumber \\
  \left.+ \frac{cos2\theta}{2}\left( \left|+2\right\rangle\left|-2\right\rangle + \left|-2\right\rangle\left|+2\right\rangle \right)         \right] 
\label{equation64b}	
\end{eqnarray}
\label{equation64}
\end{subequations} 
These equations show that a pure circularly polarized excitonic phase ($\theta=\pm\pi/4$) is perfectly stable, to first order with respect to $n_{ex}\left(a^{2D}_B \right)^2$  as stated in ref. \cite{Fernandez-Rossier}, while an elliptically polarized becomes more and more unstable when the ellipticity increases. The transfer rate $\tau^{-1}_{1,2}$  between elliptical excitons and dark excitons is proportional to $\tau^{-1}_{1,2}\propto\kappa_1 n_{\theta}cos^2(2\theta)$,where $\kappa_1$ is a constant, so it is maximum for linear excitons. The constant $\kappa_1$ can be evaluated with first order perturbation theory be of order $\kappa_1 \approx\frac{\pi}{2\hbar}K^2_1\mathcal{D}^{2D}_X\sim350cm^2s^{-1}$, where $\mathcal{D}^{2D}_X$ is the XH exciton density of state. Typical values for $\kappa_1$ are obtained from fits to the experiments, and range from $20 cm^2s^{-1}$ in samples presenting exciton localization, to $250 cm^2s^{-1}$ in homogeneous samples where localization is weak \cite{Amand-Robart,LeJeune2}. The initial collision phase leads to a decrease of the total emission intensity, at a rate the more efficient as the ellipticity is increased as seen on \ref{Figure3} (a-c). This decay stops when an equilibrium between $J=1$ and $J=2$ excitons is achieved, \textit{i.e.} when their populations become comparable. However, as the $J=1$ and $J=2$ excitons are nearly degenerated (the condition $ \Delta_0 << \Gamma$ is fulfilled, where $\Gamma$ is the exciton collision broadening, since $\Gamma$ is of the order of 1 meV \cite{Ciuti-Savona,deveaud2}), we have to consider the reverse process due to the action of $H_{exch}$ on the produced $J=2$ states, as well as interaction between $\left|E_{\theta}\right\rangle$ and $J=2$ excitons. More specifically we can write, using the (\ref{equation62}) hamiltonian:
\begin{subequations}
\begin{equation}
	H_{exch}\left(\frac{\left|+2\right\rangle \left|-2\right\rangle + \left|-2\right\rangle\left|+2\right\rangle}{\sqrt{2}}\right)
		= \frac{K_1}{A} \frac{\left|+1\right\rangle \left|-1\right\rangle + \left|-1\right\rangle\left|+1\right\rangle}{\sqrt{2}}   
	\label{equation65a}
	\end{equation}
\begin{equation}
  H_{exch}\left(\frac{\left|E_{\theta}\right\rangle\left|\pm2\right\rangle + \left|\pm2\right\rangle\left|E_{\theta}\right\rangle}{\sqrt{2}}\right)
		= \frac{K_1}{A}\frac{ \left|E_{\theta}\right\rangle\left|\pm2\right\rangle + \left|\pm2\right\rangle\left|E_{\theta}\right\rangle}{\sqrt{2}}  
	\label{equation65b}	
\end{equation}
\label{equation65}
\end{subequations} 
The second equation \ref{equation65b} shows that the secondary interactions of the $J=2$ states generated by exchange with $\left| E_{\theta}\right.\left.\right\rangle$ states does not change their polarization. However, the scattering of $J=2$ states as shown in equation (\ref{equation65a}) leads to the formation of exciton pairs made of $\left|+1\right\rangle$ and $\left|-1\right\rangle$ states, which are not coherent with the initial $\left|E_{\theta}\right.\left.\right\rangle$ states. As the emission probabilities of $\sigma^+$  and of $\sigma^-$ photons by these exciton pairs are identical, whether they dissociate or go into a bound state, the circular polarization degree of the whole optically active exciton population decays, which explains the observed results for an initial elliptical population (see fig. \ref{Figure3} (d-f)). Clearly, this process cannot occur for strictly circular excitons. This mechanism can also be enhanced in real quantum wells by additional exchange coupling term between $\left|+1 \right.\left.\right\rangle$ and $\left|-1 \right.\left.\right\rangle$ states, which arises due to coupling between heavy-hole and light-hole states (this coupling is strictly zero in pure two dimensional systems) \cite{Fernandez-Rossier}.

Finally, creating an incoherent population mixing of $\left|+1\right.\left.\right\rangle$  and $\left|-1\right.\left.\right\rangle$ states leads to the \textit{re-polarization} of the optically active excitons, due to the action of the exchange Hamiltonian. This can be derived from the equation similar to (\ref{equation65a}) with the permutation of $M= \pm2$ states by $M=\pm1$, which is also valid. This leads to the simultaneous destruction of the same number of   $\left|+1\right.\left.\right\rangle$ and $\left|-1\right.\left.\right\rangle$ states, while keeping $N_{+1}-N_{-1}$ constant, so that the circular polarization of the optically active $J=1$ states increases \cite{Amand-Robart, LeJeune2}. 

To conclude this section, let us mention that experiments of spin dynamics in the context of strong coupling of excitons with the electromagnetic field in semiconductor microcavities have been also performed and analyzed. The quasi-particle resulting from this coupling is called 2D excitonic-polariton, which is the 2D analog of Hopfield 3D polaritons \cite{Hopfield}. The exciton-polaritons present a more marked bosonic character than bare excitons, due to their photon component.  Specific aspects of 2D polaritons spin dynamics which rely on their exchange driven spin dependent scattering, such as spin dependent blue shift or parametric conversion of $\left|X\right\rangle$ to $\left|Y\right\rangle$ linearly polarized states can be found \textit{e.g.} in ref. \cite{Marie-Renucci, Kavokin-A, Shelykh, Renucci-Amand, Krizhanovskii, Solnyshkov}. Finally, let us mention that optical Spin-Hall effect has been recently observed in such microcavities \cite{Leyder}. 

\section*{Acknowledgment}
The authors are grateful to L. Lombez and A. Ballocchi for their assistance in the preparation of this manuscript.

\renewcommand \thechapter{\Roman{chapter}}
\setcounter{chapter}{1}
\setcounter{section}{0}
\setcounter{equation}{0}
\setcounter{figure}{0}
\section*{Appendix I : Excitons in quantum wells }
\label{AnnexeI}
We will limit ourselves in the following to the description of excitons in type I and type II  quantum wells, and the subsequent selection rules for optical pumping. 

\section{Excitons states in type I quantum wells}
\label{Sec:AI1}

In such quantum wells, the electron and the holes are confined in the same layer. The electron-hole wave function, in the envelope function approach, can be expressed in the basis:
\begin{eqnarray}
	\Psi_{s,m_h}(\bm r_e,\bm r_h)= F_{s,m_h}(\bm r_e,\bm r_h)u_s(\bm r_e)u_{m_h}(\bm r_h)
	\label{equationAI1}
\end{eqnarray} 
where $u_s(r_e)$ and $u_{m_h}(r_h)$ are the bulk material electron and hole Bloch functions at the Brillouin zone center $k=0$, taken as identical in the QW and in the barrier for simplicity \cite{Bastard2}, $s=\pm1/2$ and $m_h=\pm1/2,\pm3/2$; $F_{s,m_h}(\bm r_e,\bm r_h)$ is the exciton envelope function. In a bulk semiconductor of symmetry $T_d$, such as most III-V semiconductors ($GaAs$, $InAs$, $InP$… ), the functions $u_s(\bm r_e)$ transform like a spin by the symmetry operations of the crystal, so they belong to the $\Gamma_6$ representation in Koster's notations \cite{Koster}, while the $u_{m_h}(\bm r_h)$, due to spin-orbit interaction, belong to $\Gamma_8$ representation. The envelope function $F_{s,m_h}u_s(\bm r_e)u_{m_h}(\bm r_h)$ is then the solution of Schr\"odinger type equations, which take into account the coupling, at $k=0$, between the different hole-bands, the confinement potentials of the structure, and the direct Coulomb interaction between the electron and the hole. A reasonable approximation to the envelope Wannier equation for an interacting electron-hole pair in the structure at the position $(\bm r_e, \bm r_h)$ can be formulated as follows \cite{Bastard2, Greene-Bajaj}:
\begin{equation}
\mathcal{H}^{jh}_{ex}\Psi_{s,m_h}(\bm r_e,\bm r_h)=E\Psi_{s,m_h}(\bm r_e,\bm r_h)\nonumber\\
\end{equation}
\begin{equation}
\mathcal{H}^{jh}_{ex}= \frac{p_{z_e}^2}{2m_e}+V_e(z_e)+\frac{p_{z_h}^2}{2m_{jh,\parallel}}+V_h(z_h)+\frac{p_{e,\bot}^2}{2m_e}+\frac{p_{h,\bot}^2}{2m_{jh,\bot}}-\frac{e^2}{\epsilon_b\left|\bm r_e-\bm r_h\right|}
	\label{equationAI2}
\end{equation} 
This formulation means that the heavy-hole ($j=h$,$m_h=\pm3/2$) and the light-hole ($j=l$,$m_h=\pm1/2$) excitons are not coupled, which is realistic for QW with a marked 2D character, \textit{i.e.} when the exciton binding energy $E_B$ is smaller than the heavy/light hole splitting $\Delta_{lh}$. The functions $V_e(z)$ and $V_h(z)$ represent the confinement potential for the electrons and the holes respectively, which are added to the usual Coulomb attraction term $V(\bm r_e-\bm r_h)=-\frac{e^2}{\epsilon_b\left|\bm r_e-\bm r_h\right|}$ [$e^2=q^2/(4\pi\epsilon_0)$]. The holes masses take into account the anisotropy between the growth axis (\textit{Oz}), chosen as the quantization direction, and the QW plane (\textit{Oxy}). The electron conduction effective mass $m_e$ is isotropic. The hole masses ($m_0$ being the free electron mass) are given by:
\begin{subequations}
\begin{eqnarray}
  \hskip -2cm \mathrm{Heavy-holes} \,\,\, (j=h): & & \nonumber\\
 & \hskip -2cm \frac{1}{m_{hh,\parallel}}=\frac{1}{m_0}\left(\gamma_1 -2\gamma_2\right); \frac{1}{m_{hh,\bot}}=\frac{1}{m_0}\left(\gamma_1 +\gamma_2\right) &
	\label{equationAI3a}
	\end{eqnarray}
\begin{eqnarray}
   \hskip -2cm \mathrm{Light-holes} \,\,\, (j=l): \nonumber\\
& \hskip -2cm \frac{1}{m_{hh,\parallel}}=\frac{1}{m_0}\left(\gamma_1 +2\gamma_2\right); \frac{1}{m_{hh,\bot}}=\frac{1}{m_0}\left(\gamma_1 -\gamma_2\right) &
	\label{equationAI3b}	
\end{eqnarray}
\label{equationAI3}
\end{subequations} 
where the  $\Gamma_i$ are the Luttinger parameters assumed to be identical in the two materials for simplicity. Suitable boundary conditions expressing the continuity of the probability density and the current density should be added to equation \ref{equationAI2} \cite{ivchenko}. The Hamiltonian in (\ref{equationAI2}) is then rewritten in the form:
\begin{eqnarray}
\hskip -0.5 cm \mathcal{H}^{jh}_{ex}= \mathcal{H}_e+\mathcal{H}_{jh}+\mathcal{H}_G+\mathcal{H}_{rel}& & \nonumber\\
 & \hskip -5 cm =\left(\frac{p_{z_e}^2}{2m_c}+V_e(z_e)\right)+\left(\frac{p_{z_h}^2}{2m_{jh,\parallel}}+V_h(z_h)\right)+\frac{P_{\bot}^2}{2M_{jh,\bot}}+\left(\frac{p_{\bot}^2}{2\mu_{jh,\bot}}-\frac{e^2}{\epsilon_b\left|\bm r_e-\bm r_h\right|}\right)  &
	\label{equationAI4}
\end{eqnarray} 
where: $M_{jh,\bot}=m_e+m_{jh,\bot}$, $\mu_{jh,\bot}^{-1}=m^{-1}_e+m^{-1}_{jh,\bot}$ , $\bm r_{e(h)}\equiv\left(\bm r_{e(h),\bot},z_{e(h)}\right)$ and:
$ \bm R_{\bot}=\left(m_e\bm r_{e,\bot}+m_{jh,\bot}\bm r_{h,\bot}\right)/M_{jh,\bot}$, $\bm P_{\bot}\equiv\bm p_{e,\bot}+\bm p_{h,\bot}$, $\bm r_\bot\equiv\bm r_{e,\bot}-\bm r_{j,\bot}$.The background dielectric constant $\epsilon_b$ is approximated as identical in the two materials for the sake of simplicity. 

As, in the case of pseudomorphic growth, the system has the translational invariance in the plane of the quantum well, the center of mass motion is still separable. It is then convenient to look for a solution of the type:
\begin{eqnarray}
F_j(\bm r_e,\bm r_h)=\frac{1}{\sqrt{A}}e^{i\bm K_\bot \bm . \bm R_\bot}\chi_e(z_e)\chi_{jh}(z_h)G_j\left( \bm r_\bot,z_e,z_h \right)
\label{equationAI5}
\end{eqnarray} 
where $A$ is the QW area, $\bm K_\bot = \bm P_{\bot}/\hbar$, $\chi_e(z_e)$  and $\chi_{jh}(z_h)$ are respectively the single electron and hole envelope functions, which satisfy the one dimensional equations: $\mathcal{H}_e\chi_e(z)=\left(E_{c,\nu_e}-E_{c,0}\right)\chi_e(z)$ and $\mathcal{H}_{jh}\chi_{jh}(z)=\left(E_{j,\nu_h}-E_{v,0}\right)\chi_{jh}(z)$ with respective eigen-energies $E_{c,\nu_e}$and $E_{j,\nu_h}$, $ E_{c,0}$ and $E_{v,0}$ referring to the conduction and valence bands energies in $k=0$ extrema of the host bulk material. The envelope function $G$ is then the solution of the equation:
\begin{eqnarray}
& & \hskip -1.5 cm\left(\mathcal{H}_e+\mathcal{H}_{jh}+\mathcal{H}_{rel}\right)G_j(\bm r_\bot,z_e,z_h) \nonumber\\
& & =\left(E-\frac{\hbar^2K^2_\bot}{2M_\bot}-\left| E_{c,\nu_e} \right| - \left| E_{j,\nu_h}\right|- E_g \right)
G_j(\bm r_\bot,z_e,z_h)
\label{equationAI6}
\end{eqnarray} 
\begin{figure}[t]
	\centering
		\includegraphics*[width=.6\textwidth]{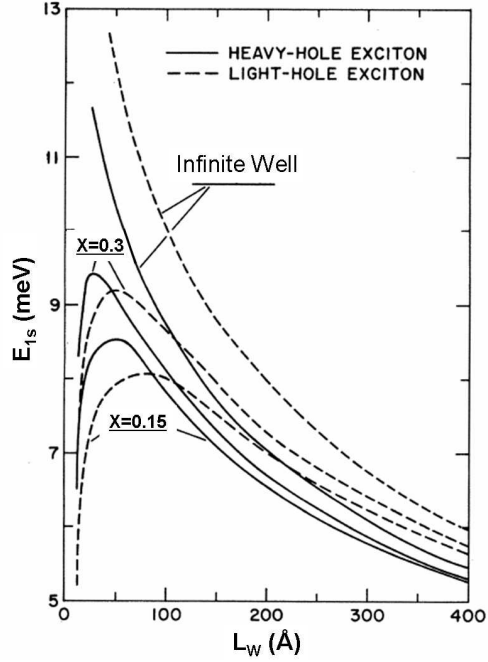}
		\caption{Variation of the binding energy of the ground state $E_{1s}$ of the heavy-hole exciton (solid lines) and the light-hole exciton (dashed lines) as a function of the $GaAs/AlGaAs$ QW thickness ($L_W$) for aluminium concentration $x=0.15$ and $x=0.3$ and finite confinement barriers. The case of infinite barriers is also shown for comparison \cite{Greene-Bajaj}.}
	\label{FigureAI1}
\end{figure}
where $E_g$ is the QW material gap. The above equation is then solved using a variational method. In type I quantum wells, taking $G_j(\bm r_\bot,z_e,z_h)=C_j\left[ 1+\alpha_j (z_e-z_h)^2e^{-\eta_jr} \right]$ with 
$r=\sqrt{\bm r^2_{\bot}+(z_e-z_h)^2}$ , $\alpha_j$,$\eta_j$ being variational parameters and $C_j$ a normalization constant, Greene et al. \cite{Greene-Bajaj} obtained the heavy and light hole binding energies as a function of the well width as displayed on fig. AI.1, in $GaAs/Al_xGa_{1-x}As$ QW structures for moles fractions $x=0.15$ and $0.3$. Clearly, the binding energy first increases when the well width decreases before reaching a maximum. In the finite well, the heavy and light holes curves cross, due to the hole mass reversal effect : as seen in equations \ref{equationAI3}, the "heavy-holes" are lighter than the "light-holes" in the directions perpendicular to the quantization axis (\textit{Oz}), \textit{i.e.} in the QW plane (\textit{xOy}). In narrow QWs of type I, it is possible to make the following approximation for the Coulomb potential: $ -\frac{e^2}{ \epsilon_b\sqrt{ \bm r^2_{\bot}+(z_e-z_h)^2}} \approx -\frac{e^2}{\epsilon_b r_{\bot}}$ [$e^2\equiv q^2/(4\pi\epsilon_0)$]. In that case, it is possible to separate completely the relative movement in the QW plane and along \textit{Oz}. Then, $G_j(\bm r_\bot,z_e,z_h)$ is the solution of the 2D equation: 
\begin{eqnarray}
\left(\frac{p^2_\bot}{2\mu_{jh}}-\frac{e^2}{\epsilon_b r_\bot} \right)\phi_j(\bm r_\bot)=E_{nl}\phi_j(\bm r_\bot)
\label{equationAI7}
\end{eqnarray} 
which have analytical solutions labelled $(n,l)$ \cite{Shinada,ivchenko}. The corresponding eigen-energies are : $E_{n,l}=E^{3D}_B/(n-1/2)^2 $, with $n = 1,2,…$and the angular momentum $l=0, 1,… |l|\leq n-1$. For instance, for the ground exciton state $(1,0)\equiv 1s$, we have: $\phi^{2D}_{1s}=\frac{1}{\sqrt{2\pi}a^{2D}_B}e^{-\frac{r_\bot}{2a^{2D}_B}}$ , with: $ a^{2D}_B\equiv a^{2D}_B/4$, the 2D Bohr radius (\textit{i.e.} the one which maximizes the probability to find an electron at a distance $r_{\bot}$ from the hole; using this definition, the equation $E_Ba_B= e^2/(2\epsilon_0)$ is valid both in 2D and 3D cases). In narrow type I quantum well, the 2D function $\phi_{j,nl}$ can be taken as a trial function of the equation \ref{equationAI6}, which is a reasonable approximation in the case of strong 2D confinement \cite{bastard}. Turning back to the full electron-hole wave function, it can finally be written as: 
\begin{eqnarray}
\Psi_{s,m_h}(\bm r_e,\bm r_h)=\chi_{c,\nu_e}(z_e)\chi_{j,\nu_h}(z_h)\frac{e^{i\bm K_\bot \bm . \bm R_{\bot}}}{\sqrt{A}}\phi^{2D}_{j,nl}(\bm r_\bot)
u_s(\bm r_e)u_{m_h}(\bm r_h)
\label{equationAI8}
\end{eqnarray} 

The function basis represented by equation \ref{equationAI8} is the usual starting point for estimating the different contribution of electron-hole exchange within the exciton.

\section{Exciton states in type II quantum well structures}
\label{Sec:AI2}

Different variational approaches have been used to describe such systems. For a $GaAs$ slab of thickness $L_W$ embedded in $AlAs$ for instance, if $L_W$ is sufficiently small, the electron is located in the $AlAs$ layer while the hole is still confined in the the $GaAs$ one. In the case where $L_W$ is still thicker enough to neglect tunnelling of the electron from the $ z > L_W/2$ to the $ z < L_W/2$ regions, the exciton wave function will have its maximum close to the interface between the two materials. Taking the exciton located nearby $z \approx L_W/2$, and in the infinite barrier approximation, one can use for instance the variational function \cite{ivchenko}:
\begin{eqnarray}
F_h(\bm r_e,\bm r_h)=\frac{1}{\sqrt{A}}e^{i\bm K_{\bot}\bm . \bm R_{\bot}}Cf(z_e)g(z_h)\chi_{h,1}(z_h)G_h(\bm r_\bot,z_e,z_h)
\label{equationAI9}
\end{eqnarray} 
where: $f(z)=(z-L_W/2)e^{-\beta_e(z-L_W/2)}Y(z-L_W/2)$, ($Y(z)$ is the Heaviside step function), $g(z)=e^{-\beta_h(L_W/2-z)}$,\\ $G_h(\bm r_\bot,z_e,z_h)=$ 
$\frac{1}{\sqrt{\pi a_{\parallel} a^2_{\bot} } }e^-{\left(r^2_{\bot}/a^2_{\bot}+(z_e-z_h)^2/a^2_{\parallel}\right)^{1/2}}$, and $C$ is a suitable normalization constant. The four variational parameters $a_{\parallel}$,$a_\bot$,$\beta_e$,$\beta_h$ determine the exciton binding energy. It can be shown that the binding energy is significantly reduced with respect to the bulk $GaAs$ one, despite the pronounced size quantization of the hole in the GaAs slab. 

\section{Optical pumping of exciton: selection rules}
\label{Sec:AI3}

The optical selection rules play a crucial part in the optical orientation experiments of excitons. The creation probability amplitude of an exciton in a state $\left|\alpha \right\rangle=\left|s,m_h;\nu_e,\nu_h,\bm K_{\bot},j,n,l\right\rangle$ by light polarized along the unit vector $\bm e$ is determined, in dipolar approximation ($K_{\bot}<< \pi /L_W$) by the matrix element:
\begin{eqnarray}
\bm e \bm .\bm r_{\alpha,\emptyset} \equiv \bm e \bm . \left\langle\alpha\left|\hat{r}\right|\emptyset \right\rangle \approx \bm e \bm .\left(\phi^{2D}_{n,l}(r=0)\right)^* \left\langle \nu_e\left|\right.\nu_h \right\rangle \left\langle s \left| \hat{\bm r}\hat{K}\right|m_h\right\rangle
\label{equationAI10}
\end{eqnarray} 
where $\left|\emptyset\right\rangle$ is the crystal fundamental state (without excitons), and $\hat{K}$ is the time reversal operator, which transforms hole states into electron valence band states. The exciton oscillator strength $f_{ex,j}$ for an optical mode of wavevector $\bm q = (\bm q_\bot,q_z)$ is then proportional to:
\begin{eqnarray}
f_{ex,j}\propto\delta_{\bm K_{\bot},\bm q_{\bot}}\frac{\hbar^2}{m^2_0 E^2_g}
\left| \left\langle \nu_e\left|\right. \nu_h\right\rangle \right|^2 
\left|\phi^{2D}_{j;n,l}(0)\right|^2 
\left| \bm e \bm . \left\langle s \right| \hat{\bm p}\hat{K}\left|m_h\right\rangle \right|^2
\label{equationAI11}	
\end{eqnarray}
where we have used the identity: $\hat{\bm p}=i m_0 \frac{E_g}{\hbar} \hat{\bm r}$.
The selection rules follow: (\textit{i}) due to in plane translational invariance, the exciton wave vector $\bm K_{\bot}$  must be the same as the projection $\bm q_{\bot}$ of the photon wave vector on the QW plane. (\textit{ii}) The QW conduction and valence states  $\nu_e$ and  $\nu_h$ must have the same parity, and (\textit{iii}) only "$ns$" 2D exciton states with $l=0$ are optically active, since the orbital angular momentum of $\hat{K}\left|m_h\right\rangle$ state is $L=1$ and $s$  states have $L=0$. (\textit{iv}) The matrix elements $\bm e \bm .  \left\langle s \right| \hat{\bm p}\hat{K}\left|m_h\right\rangle $ are the same as in the bulk material (cf intro M. Dyakonov). They express the conservation of the photon angular momentum when creating an exciton. Finally, it is clear from (\ref{equationAI11}) that the exciton oscillator strength is much lower in a type II QW than in type I, since the overlap between electron and hole is strongly reduced in the former with respect to the latter.  

%
\setcounter{chapter}{2}
\setcounter{section}{0}
\setcounter{equation}{0}
\setcounter{figure}{0}
\section*{Appendix II: Exciton fine structure in transverse magnetic field and hole spin relaxation}
\label{AnnexeII}
The observation of exciton spin or electron spin quantum beats in transient luminescence experiments can be easily explained in terms of the strength of the exchange coupling between the electron and the hole spins within the exciton.
 Choosing now the quantization axis along the transverse magnetic field $B_x \bm e_x$, we obtain new electron-hole pair basis states, namely $\left| \pm 3/2 \right. \left. \right\rangle_x \left|\pm1/2 \right. \left. \right\rangle_x$, where:
  $\left|\pm 1/2 \right. \left. \right\rangle _x = \left( \left|+1/2 \right. \left. \right\rangle \pm \left|-1/2 \right. \left. \right\rangle\right)$ , and $\left|\pm 3/2 \right. \left. \right\rangle _x= \left( \left|+3/2\right. \left.\right
   \rangle\pm\left|-3/2\right\rangle\right)$ (in case of strong confinement, the light-hole components of the heavy-hole states can be neglected). Neglecting the hole transverse g-factor ($q\approx 0$), these states are eigenstates of $\mathcal{H}_{B,\bot}\approx\omega S_x$; they are all optically active in \textit{e.g.} $\sigma^+$ polarization. The matrix of the \ref{equ13} hamiltonian then becomes, in the basis $\left\{  \left|3/2\right. \left.\right\rangle_x\left|-1/2 \right. \left. \right\rangle_x,\left|-3/2\right. \left. \right\rangle_x\left|1/2\right. \left. \right\rangle_x,\left|3/2\right. \left. \right\rangle_x\left|1/2\right. \left. \right\rangle_x,\left|-3/2\right. \left.\right\rangle_x\left|-1/2\right. \left.\right\rangle_x  \right\}$:
\begin{eqnarray}
	\mathcal{H}= 
\left[ \begin{array}{cccc}
-\frac{\hbar\omega}{2}-i\gamma_2&-\frac{\Delta_0}{2}&0&0\\
-\frac{\Delta_0}{2}&\frac{\hbar\omega}{2}-i\gamma_1&0&0\\
0&0&\frac{\hbar\omega}{2}-i\gamma_1&-\frac{\Delta_0}{2}\\	
0&0&-\frac{\Delta_0}{2}&-\frac{\hbar\omega}{2}-i\gamma_2\\ \end{array} \right]  
	\label{equationAII1}
\end{eqnarray} 
\begin{figure}[t]
	\centering
		\includegraphics*[width=0.8\textwidth]{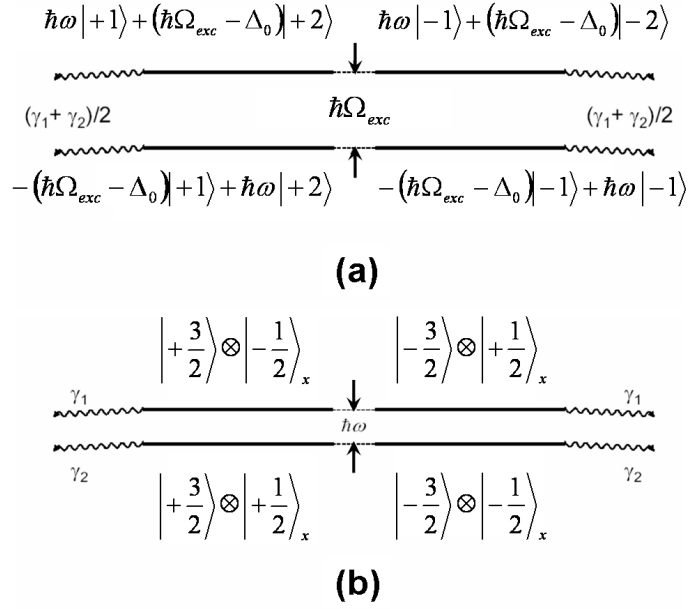}
		\caption{The effective electron-hole level scheme for: (a) strong exchange coupling between electron and hole spins; (b) weak exchange coupling between electron and hole spins. The quantization axis is taken along \textit{Oz}, so that $\left|\pm1/2\right\rangle_x =\left( \left|+1/2\right\rangle \pm \left|-1/2\right\rangle \right)/\sqrt{2}$, and it is assumed that $\hbar\omega<\Delta_0$. }
	\label{FigureAII1}
\end{figure}
where we have added the complex energy terms $ \gamma_j= \hbar / \tau_{hj} (j=1,2)$. These terms correspond to the broadening of electron-hole spin states due to the single hole spin relaxation. We neglect also here long-range exchange coupling between $\left|+1\right\rangle$ and $\left|-1\right\rangle$ exciton states. The electron-hole exchange terms $-\frac{2}{3}\delta_0\hat{J}_z\hat{S}_z $ induces thus a coupling between \begin{small}$\left|+3/2\right\rangle_x\left|\pm1/2\right\rangle_x$\end{small} and \begin{small}$\left|-3/2\right\rangle_x\left|\pm1/2\right\rangle_x$\end{small} states respectively. The eigen-energies of $\mathcal{H}$, all doubly degenerated, are easily derived, and, in the two limiting cases $\gamma_j << \delta_0 $ or $ \gamma_j >> \delta_0 $, we obtain respectively, in first order approximation: 
\begin{subequations}
\begin{eqnarray}
	\gamma_{1,2}<<\Delta_0,\,\,\, E_{\pm}\approx \pm \frac{1}{2}\hbar\Omega_{exc}-i\left( \frac{\gamma_1+\gamma_2}{2}+\frac{\gamma_1-\gamma_2}{2}\frac{\omega}{\Omega_{exc}} \right) 
	\label{equationAII2a}
	\end{eqnarray}
\begin{eqnarray}
	\hskip -4 cm \gamma_{1,2}>>\Delta_0,\,\,\, E_{\pm}\approx \pm \frac{1}{2}\hbar\omega-i\gamma_{1(2)}
	\label{equationAII2b}	
\end{eqnarray}
\label{equationAII2}
\end{subequations} 	
In the first regime \ref{equationAII2a}, the short-range exchange term $-\frac{2}{3}\Delta_0\hat{J}_z\hat{S}_z $   in the
 hamiltonian \ref{equ13} produces strong coupling between the electron and hole spins, so that electron-hole exchange
  contributes effectively the splitting $\hbar\Omega_{exc}$ between $\left|+3/2\right\rangle_x\left|-1/2\right\rangle_x$
   $\left(\left|+3/2\right\rangle_x\left|+1/2\right\rangle_x\right)$ and
    $\left|-3/2\right\rangle_x\left|+1/2\right\rangle_x$
     $\left(\left|-3/2\right\rangle_x\left|-1/2\right\rangle_x\right)$ electron-hole states. The electron-hole energy scheme is displayed in figure II.1, where we have turned back to the more convenient \textit{Oz} quantization axis. The exciton beats are then observed, till they are damped due to the MAS process (which rely on long-range electron-hole exchange). In the second case, the electron and hole spins are in the weak coupling regime, leading to the collapse of the eigen-energies, so that the electron-hole exchange does not manifests itself any more in the electron-hole energy levels. The splitting of the latter corresponds now to the Zeeman splitting $\hbar\omega$ of the electrons as if they where alone, and the electrons can be treated as independent from the holes. Electron beats corresponding to single electron Larmor precession will then be observed, since the hole spin flip does not affect the electron spin-states coherence, till they are damped due to the Dyakonov and Perel relaxation process (which rely on electron spin-orbit interaction in the conduction band).

%
\setcounter{chapter}{3}
\setcounter{section}{0}
\setcounter{equation}{0}
\setcounter{figure}{0}
\section*{Appendix III: Exciton-exciton Coulomb scattering: direct and exchange terms}
\label{AnnexeIII}
We recall now briefly the main properties of exciton-exciton Coulomb scattering in 2D systems, including exciton spin degree of freedom. We follow here the approach of C. Ciuti et al. \cite{Ciuti-Savona}, limiting ourselves to a two exciton system. Taking the 2D exciton in the 1s state envelope function (cf. Appendix I), and neglecting heavy-hole light-hole valence band mixing, it is possible to define the probability amplitude $\chi_S\left(s_e,j_h\right)=\left\langle s_e,j_h\left|\right. S\right\rangle$, where $\left|S\right\rangle$  is a linear combination of $\left|M\right\rangle$ states ($M=\pm 1,\pm 2$). For instance, $\chi_{+1}\left(s_e,j_h\right)=\delta_{s_e,-1/2}\delta_{j_h,+3/2}$. In the case of excitation by elliptical light, excitons are created in the elliptical states $\left|E_\theta\right\rangle$ (see equation \ref{equation61}), so that  $\chi_{S(\theta)}(s_e,j_h)=sin\left(\theta+\frac{\pi}{4}\right)\chi_{+1}\left(s_e,j_h\right)+cos\left(\theta+\frac{\pi}{4}\right)\chi_{-1}\left(s_e,j_h\right)$. 

The basic two exciton interaction we consider is the elastic Coulomb scattering process:
\begin{equation}
(1s,\bm K,S)+(1s,\bm K',S') \longrightarrow (1s,\bm K+\bm Q,S)+(1s,\bm K'-\bm Q,S')
\label{equationAIII2}
\end{equation} 
where the lowest $1s$ two dimensional exciton states $\left|1s,\bm K,S\right\rangle$ ($\nu_e=\nu_h=1$) can be represented by the wave function : 
\begin{eqnarray}
\left\langle \bm r_e,\bm r_h,s_e,j_h \left|\right. 1s,\bm K, S \right\rangle =\Psi_{\bm K}(\bm r_e,\bm r_h)\chi(s_e,j_h)\nonumber\\
=\frac{1}{\sqrt{A}} e^{i\bm K \bm .\bm R}\frac{1}{\sqrt{2\pi}a^{2D}_B}e^{-\frac{r}{2a^{2D}_B}} \chi(s_e,j_h)
\label{equationAIII3}
\end{eqnarray} 
Here, $\Psi_{\bm K}(\bm r_e,\bm r_h)$ is the two band envelope function of the HH exciton, $R$ and $r$ are its centre of mass and relative motion coordinates respectively, and $A$ is the quantization area. 
$\phi_{1s}(\bm r)=\frac{1}{\sqrt{2\pi}a^{2D}_B}e^{-\frac{r}{2a^{2D}_B}}$ is the relative motion wave function(c.f. appendix I) \footnote{In the strictly 2D approach we adopt here, the single particle envelope functions $\chi_{\nu_e}(z)$ and $\chi_{\nu_h}(z)$ play no significant role, and are dropped for simplicity}. Inelastic scattering channels to bound biexciton states are neglected here for simplicity. As a fact, time-resolved photoluminescence experiments on GaAs/AlGaAs QW can be described without including them. 
Neglecting here electron-hole exchange, which actually produces very small splitting ($\sim0.1 meV$ within an exciton, as seen previously c.f. \ref{sec:1}, and even less between two excitons), it is possible to build two-exciton states which are symmetrical with respect to exciton transposition (simultaneous transposition of two constituting fermions), but antisymmetric with respect to single fermion (electron or hole) transposition:    
\begin{eqnarray}
\Phi^{S,S'}_{\bm K,\bm K'}(\bm r_e,s_e,\bm r_h,j_h,\bm r'_e,s'_e,\bm r'_h,j'_h)= \nonumber\\
& & \hskip - 4.5 cm \frac{1}{\sqrt{2}} \left\{ 
\frac{1}{\sqrt{2}}\left[\Psi_{\bm K}(r_e,r_h)\chi_S(s_e,j_h)\Psi_{\bm K'}(r'_e,r'_h)\chi_{S'}(s'_e,j'_h) \right.   \right.\nonumber\\
& & \hskip - 3.2 cm \left. +\Psi_{\bm K}(r'_e,r'_h)\chi_S(s'_e,j'_h)\Psi_{\bm K'}(r_e,r_h)\chi_{S'}(s_e,j_h)\right]\nonumber\\
& & \hskip - 3.9 cm -\frac{1}{\sqrt{2}}\left[\Psi_{\bm K}(r'_e,r_h)\chi_S(s'_e,j_h)\Psi_{\bm K'}(r_e,r'_h)\chi_{S'}(s_e,j'_h)\right.      \nonumber\\
& & \hskip - 3.2 cm \left. +\Psi_{\bm K}(r_e,r'_h)\chi_S(s_e,j'_h)\Psi_{\bm K'}(r'_e,r_h)\chi_{S'}(s'_e,j_h)\right]                  \bigg\} 
\label{equationAIII4}
\end{eqnarray} 
Considering the four particle hamiltonian:
\begin{equation}
H=\frac{p^2_e}{2m_e}+\frac{p^2_h}{2m_e}+\frac{p'^2_e}{2m_e}+\frac{p'^2_h}{2m_e}+V_{int}(\bm r_e,\bm r_h,\bm r'_e,\bm r'_h)
\label{equationAIII5}
\end{equation}
where: $V_{int}(\bm r_e,\bm r_h,\bm r'_e,\bm r'_h)=-V\left(\left|\bm r_e-\bm r_h\right|\right)-V\left(\left|\bm r'_e-\bm r'_h\right|\right)
+ V\left(\left|\bm r_e-\bm r'_e\right|\right)+ V\left(\left|\bm r_h-\bm r'_h\right|\right)
- V\left(\left|\bm r_e-\bm r'_h\right|\right)- V\left(\left|\bm r'_e-\bm r_h\right|\right)$ and $V(\bm r)=e^2/(\epsilon_0 r)$, the scattering amplitude of the process \ref{equationAIII2} is given by : 
\begin{eqnarray}
H^{S_f,S'_f}_{S,S'}(\bm K,\bm K',\bm Q)=\left\langle \Phi^{S,S'}_{\bm K,\bm K'} \right|H\left|\Phi^{S_f,S'_f}_{\bm K+\bm Q,\bm K'+\bm Q}\right\rangle
\label{equationAIII6}
\end{eqnarray}
It was shown in \cite{Ciuti-Savona} that $ H^{S_f,S'_f}_{S,S'}(\bm K,\bm K',\bm Q) $ takes the form:
\begin{eqnarray}
& & H^{S_f,S'_f}_{S,S'}(\bm K,\bm K',\bm Q)= \nonumber\\
& & \left\langle S\left|\right.S_f\right\rangle \left\langle S'\left|\right.S'_f\right\rangle H_{dir}(\bm K,\bm K',\bm Q) +\left\langle S\left|\right.S'_f\right\rangle \left\langle S'\left|\right.S_f\right\rangle H^X_{exch}(\bm K,\bm K',\bm Q) \nonumber\\
& & +\mathcal{S}^e_{exch}(S,S',S_f,S'_f)H^e_{exch}(\bm K,\bm K',\bm Q)+\mathcal{S}^h_{exch}(S,S',S_f,S'_f)H^h_{exch}(\bm K,\bm K',\bm Q)\nonumber\\
\label{equationAIII7}
\end{eqnarray}
Here, $H_{dir}$ is the direct Coulomb term, which corresponds to the classical electrostatic interaction between the two excitons, $H^X_{exch}$ is the term corresponding to exciton-exciton exchange as a whole. The third and fourth terms correspond to exchange of a single electron, or single hole respectively. The factors $\mathcal{S}^e_{exch}$ and $\mathcal{S}^h_{exch}$ are given by the spin exchange sums. For instance, for electron exchange :
\begin{eqnarray}
\mathcal{S}^e_{exch}(S,S',S_f,S'_f)=\sum_{s_e,j_h,s'_e,j'_h}\chi^*_S(s_e,j_h)\chi^*_{S'}(s'_e,j'_h)\chi^*_{S_f}(s'_e,j_h)\chi^*_{S'_f}(s_e,j'_h)\nonumber\\
\label{equationAIII8}
\end{eqnarray}
A similar expression holds for $\mathcal{S}^h_{exch}$. The scattering processes are schematically represented on figure (AIII.1). 

\begin{figure}[t]
	\centering
		\includegraphics*[width=0.8\textwidth]{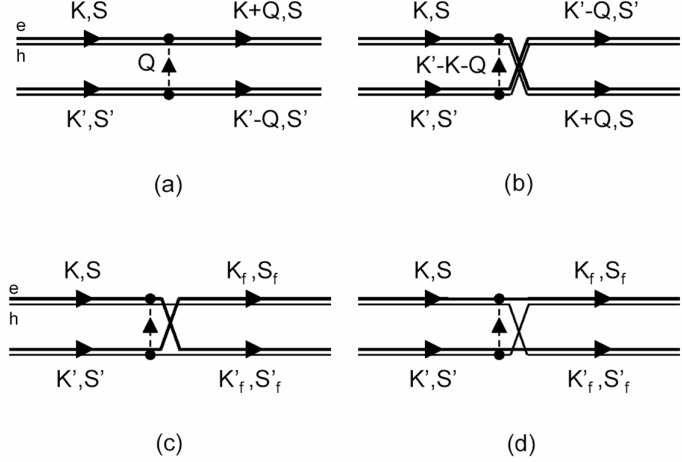}
		\caption{Scheme of the Coulomb scattering processes between two heavy-hole excitons : (a) direct process ; (b) exciton exchange as a whole; (c) electron exchange; (d) hole exchange. The electron-hole exchange processes, less efficient, has been neglected. }
	\label{FigureAIII1}
\end{figure}

The orbital part of the different scattering amplitudes in \ref{equationAIII7} has been calculated in \cite{Ciuti-Savona}. 

The direct term is the most simple, and can be formulated as:
\begin{eqnarray}
H_{dir}(\bm K,\bm K',\bm Q)=H_{dir}(\bm Q)\nonumber\\
                    & & \hskip -2 cm      =\frac{1}{A}V(Q)\left\{ \mathcal{F}\left( \left|\phi^{2D}_{1s} \right|^2 \right)(\beta_eQ)
                                                  - \mathcal{F}\left( \left|\phi^{2D}_{1s} \right|^2 \right)(\beta_hQ) \right\} \nonumber\\
        & &  \hskip -2 cm     =\frac{1}{A}\frac{2\pi e^2}{\epsilon_0Q} \left\{ \left[1+\left( a^{2D}_B \beta_eQ \right)^2\right]^{-\frac{3}{2}}    
                                                                           -\left[1+\left( a^{2D}_B \beta_hQ \right)^2\right]^{-\frac{3}{2}} \right\} \nonumber\\      
\label{equationAIII9}
\end{eqnarray}
Here, $V(Q)$ is the 2D Fourier transform of $V(r)$ and $\mathcal{F}\left(\left|\phi^{2D}_{1s}\right|^2\right)(Q)$ is the Fourier transform of the probability areal density $\left|\phi^{2D}_{1s}(\bm r)\right|^2$ of the relative motion wave function and $\beta_{e(h)}$ is the ratio defined by $\beta_{e(h)}\equiv m_{e(h\bot)}/(m_e+m_{h\bot})$. 
\begin{figure}[t]
	\centering
		\includegraphics*[width=0.8\textwidth]{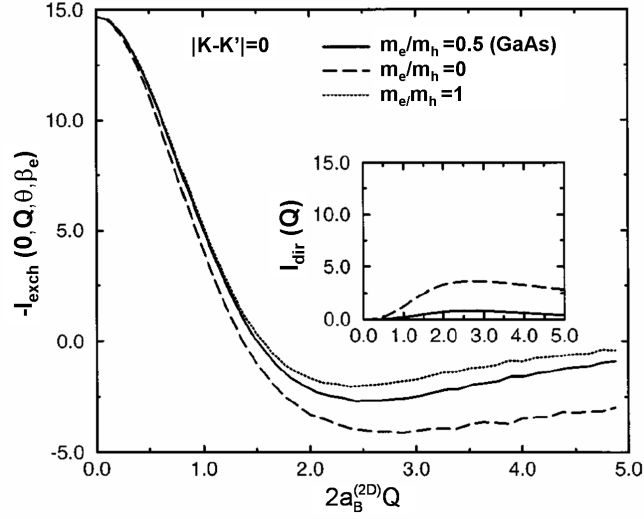}
		\caption{The dimensionless electron-electron exchange integral $-I_{exch}=H^e_{exch}\left(\left|\bm K'-\bm K \right|=0,Q,\theta,\beta_e\right)/C$ as a function of $2Qa^{2D}_B$ for three different values of $m_e/m_h$ . Solid line: $m_e /m_h=0.5$, corresponding to a GaAs quantum well. Long-dashed line: $m_e/m_h =0$. Dotted line: $m_e/m_h =1$. Inset: the dimensionless direct integral $I_{dir}(Q)=H_{dir}(Q)/C$ is shown for comparison \cite{Ciuti-Savona}.}
	\label{FigureAIII2}
\end{figure}

The exciton exchange term is linked to the direct one by the relation :
\begin{eqnarray}
H^X_{exch}(\bm K,\bm K',\bm Q)=H_{dir}(\bm K,\bm K',\bm K'-\bm K-\bm Q)=H_{dir}(\left|\bm K'-\bm K\right|,Q,\theta) \nonumber\\
\label{equationAIII10}
\end{eqnarray}
where $\theta=(\bm K'-\bm K,\bm Q)$. If $\left|\bm K'-\bm K\right|a^{2D}_B<<1$, we can, from \ref{equationAIII9}, make the approximation: $ H^X_{exch}(\bm K,\bm K',\bm Q)\approx H_{dir}(Q)$.

The single electron and hole exchange terms are more difficult to compute. It turns that they are positive real numbers, and that \cite{Ciuti-Savona}:
\begin{subequations}
\begin{equation}
   H^e_{exch}(\bm K,\bm K',\bm Q)= H^e_{exch}\left(\left|\bm K-\bm K'\right|,Q,\theta,\beta_e\right)
			\label{equationAIII11a}
	\end{equation}
\begin{equation}
	 H^h_{exch}(\bm K,\bm K',\bm Q)= H^h_{exch}\left(\left|\bm K-\bm K'\right|,Q,\theta,\beta_h\right)
	\label{equationAIII11b}	
\end{equation}
\label{equationAIII11}
\end{subequations} 	
For $\left|K-K'\right|=0$, $H^e_{exch}(0,Q,\theta,\beta_e)=H^e_{exch}(0,Q,\theta,\beta_h)$, and both become independent of $\theta$. The variations of $H^e_{exch}(0,Q,\theta,\beta_e)$ and $H_{dir}(Q)$ (normalised to the energy : $C=\left(\frac{2}{\pi}\right)^2 \frac{2e^2a^{2D}_B}{\epsilon_0A}$) are plotted on fig (AIII.2). The dependence on the mass ratio $m_e/m_h$ of the single particle exchange $H^{e(h)}_{exch}$ is not critical, contrary to the direct $H_{dir}$ or exciton exchange $H^X_{exch}$amplitudes. It turns that, for $Qa^{2D}_B=0$ the single particle exchange terms are at their maximum, while the direct one vanishes. This situation corresponds to a cold exciton gas photogenerated close to $K=0$, so that final states lie also close to $K=0$. In that case, the single particle exchange amplitude is largely dominant, and can be taken as a constant \cite{Ciuti-Savona}, the latter being close to $H^e_{exch}(0,0,\theta,\beta_e)\approx2\frac{6e^2a^{2D}_B}{\epsilon_0A}$.   

As for the spin exchange amplitude, it is more convenient to express it in terms of effective spin Hamiltonian, with the following operators: defining the heavy-hole effective spin as $\left|\pm3/2\right\rangle\equiv\left|\mp1/2\right\rangle_h$, and $\bm s_{e(h)}=\bm S_{e(h)}/\hbar$ the single spin operators, we obtain for an exciton pair $(i,j)$:
\begin{subequations}
\begin{equation}
  \mathcal{S}^e_{exch}=2\bm s^{i}_e \bm . \bm s^{j}_e + 1/2= 2 s^{i}_{e,z} s^{j}_{e,z} +  s^{i}_{e,+} s^{j}_{e,-} +s^{i}_{e,-} s^{j}_{e,+} + 1/2
  \label{equationAIII12a}
	\end{equation}
\begin{equation}
	 \mathcal{S}^h_{exch}=2\bm s^{i}_h \bm . \bm s^{j}_h + 1/2= 2 s^{i}_{h,z} s^{j}_{h,z} +  s^{i}_{h,+} s^{j}_{h,-} +s^{i}_{h,-} s^{j}_{h,+} + 1/2
	\label{equationAIII12b}	
\end{equation}
\label{equationAIII12}
\end{subequations} 	

Finally, for a cold exciton gas close to $K=0$, it is possible to write the approximate scattering Hamiltonian as : 
\begin{eqnarray}
H^{S_f,S'_f}_{S,S'}(\bm K,\bm K',\bm Q)\approx\frac{6e^2a^{2D}_B}{\epsilon_0A}4\left( \bm s^{i}_e \bm . \bm s^{j}_e + \bm s^{i}_h \bm . \bm s^{j}_h + 1/2 \right)
\label{equationAIII13}
\end{eqnarray} 






\end{document}